\documentclass[10pt]{iopart}
\usepackage{graphicx,epsfig}
\usepackage{bm}
\usepackage{iopams}
\newcommand{\ba}{\begin{eqnarray}}
\newcommand{\ea}{\end{eqnarray}}
\newcommand{\bse}{\numparts}
\newcommand{\ese}{\endnumparts}

\newcommand{\DD}{{\cal {D}}}

\newcommand{\bbq}{\begin{quote}}
\newcommand{\eeq}{\end{quote}}
\newcommand{\tbb}{t_{\textrm{\tiny{bb}}}}

\newcommand{\tcoll}{t_{\textrm{\tiny{coll}}}}

\newcommand{\CR}{{\cal{R}}}
\newcommand{\RR}{{\cal{R}}^{(3)}}

\newcommand{\T}{{}^3{\cal{T}}}

\newcommand{\EE}{{\cal{E}}}
\newcommand{\FF}{{\cal{F}}}

\newcommand{\VV}{{\cal{V}}}
\newcommand{\HH}{{\cal{H}}}
\newcommand{\HHav}{\langle{\cal{H}}\rangle}
\newcommand{\KK}{{\cal{K}}}

\newcommand{\QQ}{{\cal{Q}}}

\newcommand{\Aav}{\langle A\rangle}
\newcommand{\Bav}{\langle B\rangle}
\newcommand{\Omav}{\langle\Omega\rangle}

\newcommand{\DENL}{{\textrm{\bf{D}}}_{\textrm{\tiny{NL}}}}

\newcommand{\Da}{\delta^{(A)}}
\newcommand{\Daav}{\delta_{\textrm{\tiny{NL}}}^{(A)}}

\newcommand{\Dh}{\delta^{(\HH)}}

\newcommand{\Dk}{\delta^{(k)}}

\newcommand{\rhoav}{\langle\rho\rangle}
\newcommand{\Drho}{\delta^{(\rho)}}

\newcommand{\KKav}{\langle{\cal{K}}\rangle}
\newcommand{\DKK}{\delta^{(\KK)}}

\newcommand{\DOm}{\delta^{(\Omega)}}

\newcommand{\dd}{{\rm{d}}}
\newcommand{\Del}{{\textrm{\bf{D}}}}

\newcommand{\Sequ}{S_{\tiny{\hbox{eq}}}}

\begin{document}

\title[Weighed scalar averaging in LTB dust models, part I]{Weighed scalar averaging in LTB dust models, part I: statistical fluctuations and gravitational entropy.} 
\author{ Roberto A. Sussman}
\address{Instituto de Ciencias Nucleares, Universidad Nacional Aut\'onoma de M\'exico (ICN-UNAM),
A. P. 70--543, 04510 M\'exico D. F., M\'exico.}
\eads{$^\ddagger$\mailto{sussman@nucleares.unam.mx}}
\date{\today}
\begin{abstract} We introduce a weighed scalar average formalism (``q--average'') for the study of the theoretical properties and the dynamics of spherically symmetric Lema\^{\i}tre--Tolman--Bondi (LTB) dust models  models. The ``q--scalars'' that emerge by applying the q--averages to the density, Hubble expansion and spatial curvature (which are common to FLRW models) are directly expressible in terms of curvature and kinematic invariants and identically satisfy FLRW evolution laws without the back--reaction terms that characterize Buchert's average. The local and non--local fluctuations and perturbations with respect to the q--average convey the effects of inhomogeneity through the ratio of curvature and kinematic invariants and the magnitude of radial gradients. All curvature and kinematic proper tensors that characterize the models are expressible as irreducible algebraic expansions on the metric and 4--velocity, whose coefficients are the q--scalars and their linear and quadratic local fluctuation. All invariant contractions of these tensors are quadratic fluctuations, whose q--averages are directly and exactly related to statistical correlation moments of the density and Hubble expansion scalar.  We explore the application of this formalism to a definition of a gravitational entropy functional proposed by Hosoya {\it et al} (2004 {\it Phys. Rev. Lett.} {\bf 92} 141302--14). We show that a positive entropy production follows from a negative correlation between fluctuations of the density and Hubble scalar, providing a brief outline on its fulfillment in various LTB models and regions. While the q--average formalism is specially suited for LTB (and Szekeres) models, it may provide a valuable theoretical insight on the properties of scalar averaging in inhomogeneous spacetimes in general.                                          
\end{abstract}

\maketitle
\section{Introduction.}

The spherically symmetric Lema\^{\i}tre--Tolman--Bondi (LTB) dust models \cite{Lemaitre,TB,kras1,kras2,BKHC2009} are a very popular class of exact solutions of Einstein's equations
\footnote{This class of exact solutions was derived by Lema\^\i tre in 1930 \cite{Lemaitre}, being further investigated by Tolman and Sen in 1934, and by Bondi in 1947 (all these articles are cited in reference \cite{TB}). Articles and book reviews by Krasinski and coworkers (see \cite{kras1,kras2,BKHC2009} and references therein) refer to these solutions as ``Lema\^\i tre--Tolman'' (LT) models. However, the term ``LTB models'' has become the standard name identifying them in the literature}
. Since these models allow us to study non--linear effects in self--gravitating systems by means of analytic expressions or mathematically tractable numerical methods, they have been used in the literature in a wide variety of contexts: as  models of cosmological inhomogeneities \cite{celerier,KH1,KH2,KH3,KH4,BoKrHe,ltbstuff,focus} (see \cite{kras1,kras2,BKHC2009,celerier,focus} for reviews), as preferred test models in the effort to explain observations without resorting to dark energy \cite{celerier,obs1,obs2,kolb,GBH,alnes,bisetal,endqvist,clarkson,kras3,marranot} (see reviews in \cite{BKHC2009,marranot}), cosmic censorship \cite{lemos,joshi} and even in quantum gravity \cite{quantum}. 

The LTB models are usually described by their standard original metric variables, which are well suited for  most of their physical and cosmological applications, as can be appreciated in the abundant literature (see the book reviews \cite{kras1,kras2,BKHC2009}). However, different variables may be better suited to examine the theoretical properties of the models in a coordinate independent manner, for example, the covariant `fluid flow' scalars in the ``1+3'' formalism \cite{ellisbruni89,BDE,1plus3,zibin,dunsbyetal,LRS} whose evolution equations and spacelike constraints are equivalent to the field equations (see \cite{wainwright} for an innovative approach to the models).  Another set of alternative dynamical variables are the ``quasi--local'' scalars (to be denoted as ``q--scalars''), defined as weighed proper volume averages (``q--averages'') of covariant fluid flow scalars on spherical comoving domains. The q--scalars can be, either functionals defined on arbitrary fixed domains, or functions (``q--functions'') when considering the pointwise dependence of the average on the varying boundary of a domain. By comparing q--scalars with the non--averaged scalars we obtain fluctuations and  perturbations, which are exact, not approximated, quantities. The fluctuations and perturbations can be ``local'' when the comparison is with q--functions in a pointwise manner, or  ``non--local'' if comparing local non--averaged values with the q--average assigned to a whole domain (which in an asymptotic limit could encompass a whole time slice).  

The relevance of the present article (and its continuation, part II) follows from the fact that averaging  over inhomogeneous spacetimes has become recently an important open topic in current research in General Relativity. Evidently, as a frame dependent average acting on scalars, the q--average should be examined in reference and comparison to the similar averaging formalism of Buchert \cite{buchert}, which is widely used in the literature (see the comprehensive reviews in \cite{buchGRG,buchCQG} and \cite{zala,colpel} for an alternative covariant averaging formalism acting on proper tensors, see also \cite{avedebate} for further discussion on averaging). 

The q--average differs from  Buchert's average by the introduction of a non--trivial weight factor in the average integrals. While such a weighed proper volume average can be proposed for any spacetime, it is not evident if a procedure exists to find the right weight factor that can yield useful results for generic  spacetimes. So far, this procedure has been tried, and found successful, only for spacetimes compatible with the LTB metric (with a dust source and with nonzero pressure \cite{sussQL,suss2009,sussBR,sussIU,suss2011,sussDS1,sussDS2,RadAs,RadProfs}) and to Szekeres dust models \cite{sussbol,peel}, assuming in all cases a comoving frame.  Hence, the q--average and the results emerging from it still remain ``model dependent'' by their being specially suited for LTB and Szekeres models. On the other hand, Buchert's formalism is directly applicable to any spacetime under any time slicing without the need to find an appropriate weight factor, and thus in this respect, it may be regarded as ``model independent''.  

Since Buchert's evolution equations reveal the presence of ``back--reaction'' correlation terms that convey a significant modification of the dynamics, this formalism has a good potential for applications in Cosmology, for example, to understand the effects of non--linearity in structure formation \cite{buchstruct}, in the re--interpretation of observations \cite{buchobs}, in the possibility of explaining cosmic acceleration and dark energy \cite{buchacc}, and even in modeling dark energy sources \cite{buchdark}. As physically well motivated toy models of cosmological inhomogeneities, LTB models provide an ideal scenario to apply and test Buchert's formalism \cite{LTBave1,LTBave2,mattsson,sussBR,sussIU,suss2011}. In particular, q--functions and their local fluctuations and perturbations have already been employed as useful tools to examine the existence of back--reaction and ``effective'' acceleration in generic LTB models in the context of Buchert's formalism \cite{sussBR,sussIU,suss2011}.   

In the present article (and its continuation, part II) we extend and enhance  previous work by addressing various novel issues on the application of the q--average formalism to LTB models. In particular, besides the q--functions and local fluctuations and perturbations that were used in \cite{sussQL,suss2009,sussBR,sussIU,suss2011,sussDS1,sussDS2,RadAs,RadProfs}, we now consider also the q--scalars defined as strict average functionals and their associated non--local fluctuations and perturbations.  

The section contents of the paper are as follows: LTB models in their ``standard'' variables are described in section 2, while the q--scalars and their main properties are derived and summarized in sections 3 and 4.  We define local and non--local fluctuations and perturbations in section 5, showing that only the latter are proper statistical fluctuations. In section 6 we examine the relation of local and non--local fluctuations with all tensorial objects characteristic of the models (summarized in Appendix D). We show that all invariant contractions of proper curvature and kinematic tensors are expressible as quadratic fluctuations of the density and of the Hubble expansion scalar, while the proper tensors themselves are expressible as irreducible algebraic expansions of the metric and 4--velocity, whose coefficients are q--scalars and their local linear and quadratic fluctuations (or perturbations). In particular, the curvature tensors (Riemann, Ricci, Weyl and electric Weyl) and their contractions are entirely expressible in terms of the Ricci scalar, its q--average and its local fluctuations.  While local and non--local quadratic fluctuations of the density and Hubble scalar are different objects, their q--average is the same for every domain (see proof in Appendix C), and as a consequence, the q--average of these fluctuations are averages of invariant scalar contractions and are equal to the statistical variance and covariance (correlation) moments of the density and of the Hubble expansion scalar.       

In section 7 we provide a comparison with Buchert's averaging formalism. We show that the back--reaction terms that appear in Buchert's dynamical equations (Raychaudhuri and Friedman equations) vanish identically in the evolution equations associated with the q--scalars. Hence, the utilization of the q--average (at least as far as LTB models are concerned) leads to different implications from those of Buchert's average: instead of modifying the dynamics by the extra back--reaction terms, the q--scalars provide a more elegant and covariant description, with deeper theoretical insight, of the existing fluid flow dynamics that follows from the results of section 6: the direct (non--perturbative) relation between proper tensors and their contractions with local fluctuations and between the q--averages of quadratic fluctuations and statistical correlation moments of the density and Hubble expansion. These simple straightforward theoretically significant (and covariant) relations do not occur with Buchert's averaging (at least in its non--perturbative application to LTB models).    

As an application of the q--average formalism,  we consider in section 8 the q--averages in the context of the entropy Leibler--Kullback functional from Information Theory proposed by Hosoya, Buchert and Morita  \cite{buchGRG,entropy1,entropy2,entropy3}. We obtain the same conjecture whereby a positive entropy production from this functional follows from a negative statistical correlation between the fluctuations of the density and the Hubble expansion scalar (which is the average of an invariant quadratic contraction of the electric Weyl and shear tensors). We show by qualitative arguments (as a preliminary result) that this conjecture is fulfilled in the late time evolution of hyperbolic and elliptic models, but not in the early time evolution when the nonzero density decaying mode is dominant (non--simultaneous big bang \cite{kras2}). A summary and final comments are provided in section 9. The paper contains four brief appendices: Appendix A summarizes the standard analytic solutions of the field equations, Appendix B proves that functions of q--scalars are also q--scalars, Appendix C contains the rigorous proof that local and non--local quadratic fluctuations of the density and Hubble scalar have exactly  the same q--average for any domain, and Appendix D provides the irreducible algebraic expansions for all curvature and kinematic proper tensors characteristic of the models.

\section{The dynamics of LTB models.}\label{ltb}

The ``Lema\^\i tre--Tolman--Bondi (LTB) models'' are the well known exact solutions of Einstein's field equations characterized by the LTB metric \cite{Lemaitre,TB,kras1,kras2}:
\begin{equation} \dd s^2 = -\dd t^2 + \frac{R'{}^2}{1+2E}\,\dd r^2+R^2\left(\dd\vartheta^2+\sin^2\vartheta\,\dd\varphi^2\right),\label{ltb}\end{equation}
where $R=R(t,r)$,\, $R'=\partial R/\partial r$ and $E=E(r)$ (we have set $G=c=1$ and $r$ has length units). For a  normal geodesic comoving 4--velocity $u^a=\delta^a_0$ and a dust source $T^{ab}=\rho\,u^au^b$ with rest--mass density $\rho$ the field equations reduce to
\bse\ba  \dot R^2 = \frac{2M}{R}+2E,\label{fieldeq1}\\
 2M'= 8\pi\rho R^2R',\label{fieldeq2}\ea\ese
where $\dot R=u^aR_{,a}=\partial R/\partial t$,\,\, and $M=M(r)$ is the ``effective'' mass (the quasi--local mass function \cite{Lemaitre,podurets,MiSh,hayward}, see further discussion in \cite{suss2009}).
\footnote{This mass function is also known as the ``Misner--Sharp'' mass, however it was known before the paper by Misner and Sharp \cite{MiSh}. It was originally derived  by Lema\^\i tre \cite{Lemaitre} and rediscovered by Podurets \cite{podurets}. }

\subsection{Covariant objects.}

LTB models are characterized by the following covariant objects associated with $u^a,\,T^{ab}$ and the projection tensor $h_{ab}=g_{ab}+u_au_b$: 
\begin{itemize}
\item the rest-mass energy density given by (\ref{fieldeq2}): $\rho=u_au_bT^{ab}$: 
\item  the Hubble expansion scalar:  
\begin{equation} \fl\HH \equiv\frac{\theta}{3},\qquad\hbox{with:}\quad \theta=\tilde\nabla_a u^a=h_a^b\nabla_bu^a=\frac{2\dot R}{R}+\frac{\dot R'}{R'}, \label{HH}\end{equation}
\item the Ricci scalar of the hypersurfaces $\T[t]$ orthogonal to $u^a$ and marked by constant arbitrary $t$, whose induced metric is $h_{ab}=g_{ij}\delta_a^i\delta_b^j$ with $i,j=r,\vartheta,\varphi$:
\begin{equation} \KK \equiv \frac{\RR}{6},\qquad\hbox{with:}\quad \RR = -\frac{4(ER)'}{R^2 R'}, \label{KK}\end{equation}
\item  the shear tensor ($\sigma_{ab}=\tilde\nabla_{(a}u_{b)}-(\theta/3)h_{ab}$) and the electric Weyl tensor ($E_{ab}=u^cu^d C_{acbd}$): 
\ba \fl \sigma_{ab}=\Sigma\,\hbox{\bf{e}}_{ab},\qquad \Sigma =\hbox{\bf{e}}_{ab}\sigma^{ab}=-\frac{1}{3}\left(\frac{\dot R'}{R'}-\frac{\dot R}{R}\right),\qquad \sigma_{ab}\sigma^{ab}=6\Sigma^2,\label{Sigma1}\\
\fl E_{ab}=\EE\,\hbox{\bf{e}}_{ab},\qquad \EE =\hbox{\bf{e}}_{ab}E^{ab}=-\frac{M}{R^3}+\frac{4\pi}{3}\,\rho,\qquad\quad E_{ab}E^{ab}=6\EE^2,\label{EE1}
\ea
where $C_{abcd}$ is the Weyl tensor and the eigenvalues ($\Sigma$ and $\EE$) of $\sigma_{ab}$ and $E_{ab}$ are given in the eigenframe $\hbox{\bf{e}}_{ab}=h_{ab}-3n_a n_b$,  with $n_a=\sqrt{h_{rr}}\delta^r_a$ being a unit 4--vector orthogonal to $u^a$ and to the orbits of SO(3). 
\end{itemize}
It is evident that the dynamics of LTB models becomes completely determined once the following covariant scalars
\begin{equation}\{\rho,\,\HH,\,\Sigma,\,\EE,\,\KK\},\label{locscals}\end{equation}
have been computed, which can be done through the metric functions (see exact solutions (\ref{ellt1}) in Appendix A) or from suitable evolution equations (see equations (\ref{EV1})--(\ref{EV4}) further ahead). 

\subsection{Analytic solutions vs the ``fluid flow'' covariant approach.} 

Most of the literature dealing with LTB models and their applications (see reviews in  \cite{kras1,kras2,BKHC2009}) relies on the analytic solutions of the field equation (\ref{fieldeq1}) (see Appendix A). These solutions are given by equations (\ref{ellt1}) and fully determine the dynamics of the models, as they allow us to obtain (implicitly or parametrically) analytic expressions for quantities like $R'$ and $\dot R'$ that are needed to compute the covariant scalars (\ref{locscals}) from their forms in (\ref{fieldeq2}) and (\ref{HH})--(\ref{EE1}).  

However, the dynamics of the models can also be fully determined by finding the scalars (\ref{locscals}) through suitable evolution equations, such as those that follow from the ``1+3'' formalism of Ehlers, Ellis, Bruni, Dunsby and van Ellst~\cite{ellisbruni89,BDE,1plus3,LRS,zibin,dunsbyetal}: 
\bse\ba
\dot\HH &=&-\HH^2
-\frac{4\pi}{3}\,\rho-2\Sigma^2,\label{EV1}\\
\dot \rho &=& -3\,\rho\,\HH,\label{EV2}\\
\dot\Sigma &=& -2\,\HH\,\Sigma+\Sigma^2-\EE,
\label{EV3}\\
 \dot\EE &=& -4\pi\,\rho\Sigma
-3\,\EE \left(\HH+\Sigma\right),\label{EV4}\ea\ese
together with the spacelike and Hamiltonian constraints
\begin{equation} \left(\Sigma+\HH\right)'+3\,\Sigma\,\frac{R'}{R}=0,\qquad \left(\frac{4\pi}{3}\,\rho
+\EE\right)'+3\,\EE\,\frac{R'}{R}=0,\label{constr} \end{equation}
\begin{equation}\HH^2 = \frac{8\pi}{3}\, \rho
-\KK+\Sigma^2.\label{Hamconstr}\end{equation}
Solving this system of partial differential equations is equivalent to working out the models through the field equations (\ref{fieldeq1})--(\ref{fieldeq2}) and the solutions (\ref{ellt1}). However, (\ref{locscals}) is not the only nor the most convenient `fluid flow' scalar representation to handle the models.

\section{The q--scalars.}\label{qscals}

As shown in \cite{sussQL,suss2009,RadAs,RadProfs}, an alternative coordinate independent  ``fluid flow'' representation for the LTB models is provided by the ``quasi--local'' or q--scalars and their fluctuations and perturbations. We provide below a rigorous definition of these variables.
\footnote{See \cite{sussQL,suss2009} for a discussion on the relation between the definition of q--scalars (q--averages and q--functions in (\ref{qave}) and (\ref{qfuncts})) to the quasi--local or Misner--Sharp mass when $A=\rho$. The notation and units here are different from those of previous articles: q--scalars were denoted as $A_*$ in \cite{sussQL,suss2009}, while the symbol $m$ was used for $(4/3)\pi\rho$ and $k$ for $\KK$ in \cite{sussIU,sussBR,suss2011,RadAs,RadProfs}. The notation used in (\ref{qave}) expresses the domain dependence of the q--average by the symbol ``$[r_b]$'' that univocally identifies any domain $\DD[r_b]$ by its boundary:  the comoving 2--sphere marked by constant arbitrary $r=r_b$. We use the subindex ${}_q$ to distinguish the q--average from the standard proper volume average functional of Buchert's formalism. Both averages coincide only for parabolic models (see first class of solutions in (\ref{ellt1})) for which $\FF=1,\,E=0$ and self--similar LTB models for which $\FF\ne 1,\,E=$ const. (see \cite{suss2011}).} 

\subsection{The q--averages.} 

Consider an arbitrary complete 
\footnote{See \cite{RadProfs,suss2011} for the case of non--complete slices intersecting a curvature singularity. We assume absence of shell crossing singularities and do not consider time slices $\T[t]$ with: (i) ``closed'' topology (homeomorphic to $\mathbb{S}^3$) and (ii) lacking symmetry centers. Hence we assume that $R'>0$ holds for all $r$.}
time slice $\T[t]$ of an LTB model with metric (\ref{ltb}) whose proper volume element is 
\begin{equation} \dd\VV_p =\sqrt{\hbox{det}(h_{ab})}\,\dd^3 x= \frac{R^2 R'\sin\vartheta}{\FF}\,\dd r\,\dd \vartheta\,\dd\varphi, \label{dV}\end{equation}
where (from (\ref{fieldeq1}))
\begin{equation}\FF \equiv \sqrt{1+2E} = \left[\dot R^2 +\left(1-\frac{2M}{R}\right)\right]^{1/2}.\label{FF}\end{equation} 
Every $\T[t]$ can be foliated by the class of 2--spheres that are the boundaries of compact concentric spherical comoving domains diffeomorphic to $\DD[r_b]=\chi[r_b]\times \mathbb{S}^2\subset \T[t]$, where  $\mathbb{S}^2$ is the unit 2--sphere and the semi--closed real interval $\chi[r_b]=\{\bar r\,|\,0\leq \bar r< r_b\}$ is a segment of a radial ray with $\bar r=0$ marking the worldline of the symmetry center (fixed point of the rotation group SO(3)). Following \cite{suss2011}, we define
\begin{quote}
{\bf The q--average.} Let $X(\DD[r_b])$ denote the set of all smooth integrable LTB scalar functions in a domain $\DD[r_b]$ with $r_b$ fixed but arbitrary marking its boundary. The q--average of a scalar $A\in X(\DD[r_b])$ is the linear functional $\langle\;\;\rangle_q[r_b]: X(\DD_[r_b])\to \mathbb{R}$ such that:
\begin{equation}
\fl  A \mapsto \Aav_q[r_b]=\frac{\int_{\DD[r_b]}{A\,\FF\,\dd\VV_p}}{\int_{\DD[r_b]}{\FF\,\dd\VV_p}}=\frac{\int_0^{r_b}{A(\bar r) R^2(\bar r) R'(\bar r)\,\dd\bar r}}{\int_0^{r_b}{R^2(\bar r) R'(\bar r)\,\dd\bar r}},\label{qave}\end{equation}
where the ``weight factor'' $\FF$ is given by (\ref{FF}).
\end{quote} 

\noindent
By applying (\ref{qave}) to (\ref{fieldeq2}), (\ref{HH}) and (\ref{KK}) it is straightforward to evaluate in an arbitrary fixed domain $\DD[r_b]$ the q--average of those covariant scalars $A=\rho,\,\HH,\,\KK$ that are common with a FLRW cosmology:
\bse\ba \frac{4\pi}{3}\rhoav_q[r_b] = \frac{4\pi}{3} \frac{\int_0^{r_b}{\rho\,R^2\,R'\dd\bar r}}{\int_0^{r_b}{R^2\,R'\dd\bar r}}= \frac{2M_b}{R_b^3},\label{rhoav}\\
\HHav_q[r_b]=\frac{\int_0^{r_b}{\HH\,R^2\,R'\dd\bar r}}{\int_0^{r_b}{R^2\,R'\dd\bar r}}= \frac{\dot R_b}{R_b},\label{HHav}\\
\KKav_q[r_b]=\frac{\int_0^{r_b}{\KK\,R^2\,R'\dd\bar r}}{\int_0^{r_b}{R^2\,R'\dd\bar r}}= -\frac{2E_b}{R_b^2}=\frac{1-\FF_b^2}{R_b^2},\label{KKav}
\ea\ese 
where $R_b=R(t,r_b),\, M_b=M(r_b),\, E_b=E(r_b)$ (subindex ${}_b$ will denote evaluation at $r=r_b$), and we simply applied the definition (\ref{qave}) to (\ref{fieldeq2}), (\ref{HH}) and (\ref{KK}). The q--averages (\ref{rhoav})--(\ref{KKav}) transform  the field equation (\ref{fieldeq1}) for every $\DD[r_0]$ into
\begin{equation}\HHav_q^2[r_b]= \frac{8\pi}{3}\rhoav_q[r_b]-\KKav_q[r_b],\label{aveFriedman}\end{equation}
which is formally identical with the Friedman equation for a dust FLRW model given in terms of the q--averages of the equivalent LTB scalars. 

It is important to remark that the weight factor (\ref{FF}) is an invariant quantity, since $R$ and $M$ are invariant scalars in spherically symmetric spacetimes \cite{hayward}. In fact, in the Newtonian limit ($2M/R\ll 1$) this factor reduces to the total binding energy of comoving dust layers, while in the special relativity limit it becomes the ``gamma factor'' \cite{MiSh}.

\subsection{The q--functions.}

Since $r=r_b$ in (\ref{qave}) and (\ref{rhoav})--(\ref{KKav}) is arbitrary, the definition of the q--average functional leads in a natural way to local functions with this corresponding rule.  We define then:
\begin{quote}
{\bf The q--functions.} For every $\T[t]$ foliated by concentric domains $\DD[r]$ with $r\geq 0$,  the q--function associated with any $A\in X(\DD[r])$ is the real valued function $A_q: \T[t]\to \mathbb{R}$ such that
\begin{equation}\qquad A_q(r)=\Aav_q[r],\label{qfuncts}\end{equation}   
holds for every $r\geq 0$. Strictly speaking, the q--functions are q--averages that depend on a varying  domain boundary.
\end{quote}
Evidently,  (\ref{rhoav})--(\ref{KKav}) and (\ref{qfuncts}) imply that the q--functions associated with $A=\rho,\,\HH,\,\KK$ are
\ba \frac{4\pi}{3}\rho_q = \frac{M}{R^3},\qquad
\HH_q = \frac{\dot R}{R},\qquad
\KK_q = \frac{1-\FF^2}{R^2},\label{rhoHKq}\\
\HH_q^2 = \frac{8\pi}{3}\rho_q - \KK_q,\label{qFriedman}
\ea
where all functions above (save those inside the integral sign) are real valued functions that depend on the upper bound of the integration range $0\leq \bar r \leq r$ for arbitrary $r$ (or the domain boundary for arbitrary domains).  

\subsection{Functionals vs functions.} 

The difference between the average functional $\Aav_q[r_b]$ and its associated q--function $A_q( r)$ is subtle but important (see figures 1a and 1b of references \cite{sussBR,sussIU,suss2011}). For every fixed arbitrary domain $\DD[r_b] \subset \T[t]$ they are identical at the boundary $r=r_b$, but they differ for all points $r\ne r_b$ because $\Aav_q[r_b]$ has been assigned by (\ref{qave}) (as a functional) to the whole of  $\DD[r_b]$ and must be considered as a constant for points $r< r_b$ inside this domain in each time slice $\T[t]$, while $A_q( r)$ (as a function) varies smoothly along these inner points (and also along outer points $r>r_b$). As a consequence, the $\Aav_q[r_b]$ and $A_q( r)$ satisfy the same local derivation rules (\ref{rder})--(\ref{intparts}) at the boundary of each generic domain (see next section), but behave differently when integrated along any fixed domain in the range $0\leq \bar r\leq r_b$. Hence:
\begin{equation}\fl \langle\,\Aav_q[r_b]\,\rangle_q[r_b]=\Aav_q[r_b]\qquad \hbox{but}\qquad \langle\,A_q( r)\,\rangle_q[r_b]\ne A_{qb}=A_q( r_b),\label{aveave}\end{equation}
holds for every scalar and every domain. The difference between functionals and functions is clearly illustrated in figure 1 of \cite{sussIU} and figure 1 of \cite{suss2011}.

\subsection{Notation.}  

We will adopt henceforth the following conventions in order to simplify the notation:
\begin{itemize}
\item The time dependence of functionals and functions will be omitted unless it is needed to avoid ambiguities. The symbol $\bar r$ will be used as a dummy variable inside integral signs.
\item Unless stated otherwise, radial dependence given in terms of $r$ ({\it i.e.} ``$M( r)$'' or ``$\rho_q( r)$'' or ``$\rhoav_q[r]$'') or altogether omitted ({\it i.e.} ``$M$'' or ``$\rho_q$'' or ``$\rhoav_q$'') will be understood as local point--wise dependence like q--functions, whereas radial dependence in terms of $r_b$ will denote functionals that correspond to a fixed arbitrary domain $\DD[r_b]$ and functions that equate to them (as in the right hand sides of (\ref{rhoav})--(\ref{KKav})). 
\end{itemize} 

\section{Properties of the q--scalars.}


\subsection{Time and radial derivatives.}

It is straightforward to show that the following commutating rule holds for every scalar $A$
\begin{equation}  \frac{\partial}{\partial t} \Aav_q-\left\langle \frac{\partial A}{\partial t} \right\rangle_q= \Aav\dot{}_q-\langle \dot A\rangle_q =3\langle \HH\,A\rangle_q -3\HHav_q \Aav_q,\label{tder}\end{equation} 
where we used (\ref{HHav}) and expressed (\ref{HH}) as $3\HH = \theta = (R^2R')\dot{}/(R^2R')$. The following useful relations are readily obtained: 
\ba 
 \Aav'_q &=& \frac{\partial}{\partial r} \Aav_q=\frac{3R'}{R}\,\left[\,A-\Aav_q\,\right],\label{rder}\\
 \Aav_q &=&  A(r )-\frac{1}{R^3( r)}\int_0^{r}{A'(\bar r)\,R^3(\bar r)\,\dd\bar r},\label{intparts}\ea
where (\ref{intparts}) follows by integrating (\ref{qave}) by parts. The same properties hold identically for the q--functions $A_q$.

\subsection{Functions of q--scalars.}

As we prove in Appendix B (see also Appendix C of \cite{sussbol}), all scalars expressible as functions of q--scalars are themselves q--scalars. As an   example, consider the Omega factor (analogous to a FLRW Omega factor) defined as a q--function $\Omega_q=U(\rho_q,\HH_q)$ or $\Omega_q=U(\rho_q,\KK_q)$ by:
\begin{equation} \fl\Omega_q = \frac{8\pi\rho_q}{3\HH_q^2} = \frac{(8\pi/3)\rho_q}{(8\pi/3)\rho_q-\KK_q},\qquad \Omega_q - 1 = \frac{\KK_q}{\HH_q^2}=\frac{\KK_q}{(8\pi/3)\rho_q-\KK_q},\label{Omdef}\end{equation}
whose associated local scalar is given by (\ref{eqZ}) as
\begin{equation} \hbox{``}\Omega\hbox{''} = \Omega_q\left[1+\frac{\rho-\rho_q}{\rho_q}-\frac{2(\HH-\HH_q)}{\HH_q}\right],\label{Omloc} \end{equation}
where we used the quotation marks to emphasize that $\Omega \ne 8\pi\rho/(3\HH^2)=U(\rho,\HH)$. Notice that (\ref{Omloc}) can be expressed as $\Omega=\Omega_q(1+\DOm)$ with $\DOm$ given by (\ref{DOmdef}).   


\subsection{The q--scalars satisfy FLRW time evolution.}

It is straightforward to show that the q--scalars (\ref{rhoav})--(\ref{KKav}) and (\ref{Omdef}) (whether evaluated as q--functions or as functionals in fixed domains $\DD[r_b]$) satisfy FLRW evolution laws:
\footnote{These evolution laws also hold for the q--functions $A_q$. For the functionals $\Aav_q[r_b]$ the derivatives involved are $\Aav\dot{}_q[r_b]$, which can be evaluated (at $r=r_b$) either directly from (\ref{rhoav})--(\ref{KKav}), or with the commutation rule (\ref{tder}) and the forms of the local (non--averaged) scalars in (\ref{fieldeq2}), (\ref{HH}), (\ref{KK}) and (\ref{Omloc}). If using (\ref{tder}) for computing $\HHav\dot{}_q[r_b]$ and $\langle\Omega\rangle\dot{}_q[r_b]$ we also need to use the identities (\ref{var}) and (\ref{cov}) that are proved in Appendix C.}
\bse\ba\rhoav\dot{}_q &=& -3 \HHav_q\rhoav_q,\qquad  \KKav\dot{}_q = -2\HHav_q\KKav_q\label{FLRW1}\\
\HHav\dot{}_q &=& -\HHav_q^2-\frac{4\pi}{3}\rhoav_q,\qquad  \Omav\dot{}_q = \Omav_q\left(\Omav_q-1\right),\label{FLRW2}\ea\ese
which are identical to the evolution equations satisfied by the equivalent covariant scalars $\tilde\rho,\,\tilde\HH,\,\tilde\KK,\,\tilde\Omega$ of a FLRW dust model (a tilde will denote henceforth FLRW objects). This fact evidently singles out the q--scalars as LTB scalars that behave as FLRW scalars (in the sense that they comply with FLRW dynamics). We comment further on this issue in part II.

\section{Fluctuations and perturbations of q--scalars.}

\subsection{Local fluctuations and perturbations.}

If $A$ and $A_q=\Aav_q$ are both evaluated as real valued functions on the same arbitrary value $r$ that denotes a varying boundary of concentric domains $\DD[r]$ for $r\geq 0$, then a local fluctuation can be defined at each $r$ by the simple pointwise comparison: 
\begin{equation} \Del(A)( r) = A( r) - A_q( r) = A( r)-\Aav_q[r].\label{locfluc}\end{equation} 
By normalizing $\Del(A)$ with $A_q$ we obtain the useful dimensionless relative local fluctuations that will be called local ``perturbations'':
\begin{equation} \Da(r) = \frac{\Del(A)( r)}{A_q( r)} =\frac{A(r)-A_q(r)}{A_q( r)},\label{Dadef}\end{equation}   
which comply (from (\ref{rder}) and (\ref{intparts})) with the following useful relation with  radial gradients of $A_q$ and $A$ (also valid for the $\Aav_q$): 
\begin{equation} \Da = \frac{A'_q/A_q}{3R'/R} = \frac{1}{A_q(r)\,R^3(r)}\int_0^r{A'(\bar r)\,R^3(\bar r)\,\dd\bar r},\label{Dagrad}\end{equation}
that lead, using (\ref{rder}), (\ref{intparts}) and (\ref{eqZ}), to the following linear algebraic relations among the $\Da$:
\ba 
2\Dh= \Omega_q\,\Drho +\left[1-\Omega_q\right]\Dk,\label{Dhdef} \\
\DOm = \Drho-2\Dh=\left(1-\Omega_q\right)\left(\Drho-\DKK\right),\label{DOmdef}  \ea
where $\Omega_q$ is given by (\ref{Omdef}) and $\DOm$ above is consistent with $\Omega=\Omega_q(1+\DOm)$ in (\ref{Omloc}).

\subsection{Non--local fluctuations and perturbations.}\label{contrast}

As opposed to local fluctuations that compare $A$ with $A_q=\Aav_q$ at the same $r$, we can define for every fixed domain $\DD[r_b]$ non--local fluctuations
\begin{equation} \DENL (A)(r,r_b)=A( r) - \Aav_q[r_b],\label{nonlocfluc} \end{equation}
that compare local values $A( r)$ inside the domain with the q--average (functional) of $A$, which is a non--local quantity assigned to the whole domain (notice that at every $\T[t]$ the value $\Aav_q[r_b]$ is effectively a constant for all $r<r_b$ and a function of $t$ for varying $\T[t]$). Non--local perturbations are readily defined by
\begin{equation}\fl \Daav(r,r_b)=\frac{\DENL(A)(r,r_b)}{\Aav_q[r_b]} =\frac{A( r)-\Aav_q[r_b]}{\Aav_q[r_b]},\qquad 0\leq r< r_b,\quad \Aav_q[r_b]\ne 0,\label{Danl}\end{equation}
and, evidently, do not comply with (\ref{Dagrad}) and the properties that follow thereof (notice that $\partial/\partial r [\Daav(r,r_b)]=A'/\Aav_q[r_b]$).

\subsection{Statistical fluctuations.}

Since evaluating a q--average for a fixed arbitrary $\DD[r_b]$ involves integration through the range $0\leq \bar r\leq r_b$, and the functional $\Aav_q[r_b]$ is effectively constant in this range at every $\T[t]$, then, following (\ref{aveave}), the q--average of local and non--local linear fluctuations is different:
\begin{equation} \langle \Del(A)\rangle_q[r_b]\ne 0\quad \hbox{but}\quad  \langle \DENL(A)\rangle_q[r_b]=0,\end{equation}
which means that the $\DENL(A)$ are statistical fluctuations, but the $\Del(A)$ are not.  

However,  the average of quadratic combinations of local fluctuations of $\rho$ and $\HH$ coincides with the average of the same combinations of non--local fluctuations.  As we prove in Appendix C (see also Appendix B of \cite{suss2011} for similar results in Buchert's averages), the following results are valid for $A,\,B=\rho,\,\HH$ and for any domain $\DD[r_b]$:\\  
\ba\fl \langle\,[\Del(A)]^2\,\rangle_q=\langle\,[\DENL(A)]^2\,\rangle_q=\langle A^2\rangle_q - \Aav_q^2=\hbox{{\bf Var}}_q(A),\label{var}\\
\fl  \langle\,\Del(A)\,\Del(B)\,\rangle_q=\langle\,\DENL(A)\,\DENL(B)\,\rangle_q
=\langle AB\rangle_q - \Aav_q\,\Bav_q=\hbox{{\bf Cov}}_q(A,B),\label{cov}\ea 
where we omitted the $[r_b]$ symbol to simplify notation ($A,\,A_q$ inside inside the $\langle\hskip 0.3cm\rangle_q$ brackets depend on $\bar r \leq r_b$), whereas $\hbox{{\bf Var}}_q(A)$ and $\hbox{{\bf Cov}}_q(A)$ are the quadratic variance and covariance (correlation) moments associated with $\Aav_q[r_b]$. 

\section{Relation of q--scalars and their fluctuations with curvature and kinematic invariants and tensors.}

As shown in Appendix D, all proper curvature and kinematic tensors characteristic of LTB models are expressible in terms of algebraic combinations of $g_{ab},\, u^a$ and the eigenframe $\hbox{\bf{e}}_{ab}$ defined below (\ref{Sigma1}) and (\ref{EE1}), with their scalar coefficients given entirely by four scalar invariants: the Ricci scalar ($\CR$), the Newman--Penrose conformal invariant ($\Psi_2=-\EE$), the Hubble expansion $\HH=\theta/3$ and the eigenvalue of the shear tensor, $\Sigma$ (see (\ref{HH}) and (\ref{Sigma1})). As a consequence, the q--scalars $\rho_q,\,\HH_q$ are coordinate independent quantities, as they are directly related to these four scalar invariants:
\footnote{The covariant nature of $\rho_q,\,\HH_q,\,\KK_q$ and their corresponding functionals is also evident from (\ref{FF}), (\ref{rhoav})--(\ref{KKav}) and (\ref{rhoHKq}), since $M$ and $R$ are scalar invariants in spherically symmetric spacetimes \cite{hayward}.}
\begin{equation} \frac{4\pi}{3}\rho_q=\frac{1}{6}\CR-\Psi_2,\qquad 
\HH_q = \HH+\Sigma,\label{invq}\end{equation}
where we used (\ref{fieldeq1}), (\ref{HH})--(\ref{EE1}), (\ref{rhoav})--(\ref{aveFriedman}) and/or (\ref{rhoHKq})--(\ref{qFriedman}), and the fact that $\CR=8\pi\rho$.  We remark that these relations hold for their associated q--averages $\rhoav_q,\,\HHav_q,\,\KKav_q$ as functions of a varying domain boundary $r$. The expressions for other q--scalars, $\KK_q$ and $\Omega_q$, in terms of $\CR,\,\Psi_2,\,\Sigma,\,\HH$ can be found by eliminating these q--scalars from  the constraints (\ref{qFriedman}) and (\ref{Omdef}) and substituting (\ref{invq}).  

It is easy to show that the eigenvalues of the shear and electric Weyl tensors are local linear fluctuations over the q--averages (or q--functions) of $\HH$ and $\rho$:
\begin{equation} \fl \Sigma = -\Del(\HH) =  -\left[\,\HH-\HH_q\,\right]\qquad \EE=-\frac{4\pi}{3}\Del(\rho)=  -\frac{4\pi}{3}\left[\,\rho-\rho_q\,\right],\label{SigE2} \end{equation}
where we applied (\ref{rhoHKq}) to (\ref{HH}), (\ref{Sigma1}) and (\ref{EE1}). Considering that $\Psi_2=-\EE$ and using (\ref{Weyl2}), the curvature and kinematic proper tensors (see Appendix D) are expressible in terms of $\rho,\,\HH$ and their fluctuations $\Del(\rho),\,\Del(\HH)$:
\bse\ba \fl \CR^{ab}_{cd}=\frac{8\pi}{3}\rho\left(3\delta^{[a}_{[c}\delta^{b]}_{d]}+6\delta^{[a}_{[c}u^{b]}u_{d]}-\delta^a_{[c}\delta^b_{d]}\right)-\frac{4\pi}{3}\Del(\rho)\,\left(h^{[a}_{[c}-3u_{[c}u^{[a}\right)\,\hbox{\bf{e}}^{b]}_{d]},\label{propten1a}\\
\fl \CR^a_b = 4\pi\rho\left(h^a_b+u^a u_b\right),\label{propten2a}\\
\fl E_{ab}=-\frac{4\pi}{3}\Del(\rho)\,\hbox{\bf{e}}_{ab},
\qquad  C^{ab}_{cd}=-\frac{4\pi}{3}\Del(\rho)\,\left(h^{[a}_{[c}-3u_{[c}u^{[a}\right)\,\hbox{\bf{e}}^{b]}_{d]},\label{propten3a}\\
\fl \sigma_{ab}=-\Del(\HH)\,\hbox{\bf{e}}_{ab},\qquad \HH_{ab}=\HH h_{ab}-\Del(\HH)\,\hbox{\bf{e}}_{ab},\label{propten4a} 
\ea\ese
while their scalar contractions take the form:
\bse\ba \fl \CR_{abcd}\CR^{abcd}=\frac{256\pi^2}{3}\left([\Del(\rho)]^2+\frac{5}{4}\rho^2\right),\qquad \CR_{ab}\CR^{ab} = 64\pi^2\rho^2,\label{contr1a}\\
\fl C_{abcd}C^{abcd} =\frac{256\pi^2}{3}[\Del(\rho)]^2 = 8 E_{ab}E^{ab},\label{contr2a}\\
\fl \sigma_{ab}\sigma^{ab}= 6[\Del(\HH)]^2, \qquad \sigma_{ab}E^{ab} = \frac{4\pi}{3}\Del(\rho)\Del(\HH),\label{contr3a}\ea\ese
which depend exclusively on $\rho,\,\HH$ and their fluctuations. Considering that we have $A=A_q(1+\Da)$ and $\Del(A)=A_q\Da$ for all $A$, the tensors (\ref{propten1a})--(\ref{propten4a}) and their contractions (\ref{contr1a})--(\ref{contr3a}) can be entirely given in terms of $\rho_q,\,\HH_q$ and their local perturbations:
\bse\ba \fl E^a_b=-\frac{4\pi}{3}\rho_q\Drho\,\hbox{\bf{e}}^a_b,\qquad  \CR^a_b = 4\pi\rho_q(1+\Drho)\left(h^a_b+u^a u_b\right),\label{propten1b}\\
\fl C^{ab}_{cd}=-\frac{4\pi}{3}\rho_q\Drho\,\left(h^{[a}_{[c}-3u_{[c}u^{[a}\right)\,\hbox{\bf{e}}^{b]}_{d]},\label{propten2b}\\
\fl \CR^{ab}_{cd}=\frac{8\pi}{3}\rho_q(1+\Drho)\left(3\delta^{[a}_{[c}\delta^{b]}_{d]}+6\delta^{[a}_{[c}u^{b]}u_{d]}-\delta^a_{[c}\delta^b_{d]}\right)-\frac{4\pi}{3}\rho_q\Drho\,\left(h^{[a}_{[c}-3u_{[c}u^{[a}\right)\,\hbox{\bf{e}}^{b]}_{d]},\nonumber\\\label{propten3b}\\
\fl \sigma^a_b=-\HH_q\Dh\,\hbox{\bf{e}}^a_b,\qquad \HH^a_b=\HH_q\left[h^a_b+(h^a_b-\hbox{\bf{e}}^a_b)\,\Dh\right],\label{propten4b}\\
\fl \CR_{abcd} \CR^{abcd} = \frac{256\pi^2}{3}\rho_q^2\left([\Drho]^2+\frac{5}{4}[1+\Drho]^2\right),\label{propten5b}\\
\fl C_{abcd}C^{abcd} = \frac{256\pi^2}{3} [\rho_q\,\Drho]^2 = 8 E_{ab}E^{ab},\label{propten6b}\\
\fl \sigma_{ab}\sigma^{ab} = 6 [\HH_q\,\Dh]^2,\qquad \HH_{ab}\HH^{ab} = \HH_q^2\left[1+(2+3\Dh)\,\Dh\right],\label{propten7b}
\ea\ese
which shows that the variables $\{\rho_q,\,\HH_q,\,\Drho,\,\Dh\}$ should provide a full dynamical representation for the models, so that $\Drho=\Dh=0$ indicates a FLRW limit characterized by the vanishing of the shear tensor and Weyl part of the curvature tensors. Notice that all these tensors can also be expressed in terms of other q--scalars ($\KK_q,\,\Omega_q$) and their local perturbations ($\DKK,\,\DOm$), as the latter can be obtained from $\rho_q,\,\HH_q,\,\Drho,\,\Dh$ by means of the algebraic constraints (\ref{qFriedman}), (\ref{Omdef}), (\ref{Dhdef}) and (\ref{DOmdef}). 

Bearing in mind that $\CR_q=8\pi\rho_q$, we can express the conformal invariant $\Psi_2$ as a local fluctuation of the Ricci scalar: $\Psi_2=(1/6)\Del(\CR)=(4\pi/3)\,\Del(\rho)$, and thus express curvature tensors (the electric Weyl, Weyl and Riemann) and their contractions in terms of the the Ricci scalar and its fluctuations:
\bse\ba \fl E_{ab}=-\frac{1}{6}\Del(\CR)\,\hbox{\bf{e}}_{ab},\qquad C^{ab}_{cd}=-\frac{1}{6}\Del(\CR)\,\left(h^{[a}_{[c}-3u_{[c}u^{[a}\right)\,\hbox{\bf{e}}^{b]}_{d]},\label{ECricci}\\
\fl \CR^{ab}_{cd}=\frac{1}{3}\CR\left(3\delta^{[a}_{[c}\delta^{b]}_{d]}+6\delta^{[a}_{[c}u^{b]}u_{d]}-\delta^a_{[c}\delta^b_{d]}\right)-\frac{1}{6}\Del(\CR)\,\left(h^{[a}_{[c}-3u_{[c}u^{[a}\right)\,\hbox{\bf{e}}^{b]}_{d]},\\
\fl \CR_{abcd}\CR^{abcd} = \frac{4}{3}\left([\Del(\CR)]^2 +\frac{5}{4}\CR^2\right),\qquad C_{abcd}C^{abcd} =  \frac{4}{3}\,[\Del(\CR)]^2=8E_{ab}E^{ab},\ea\ese
where we used (\ref{ECricci}) and (\ref{Rie2}). These expressions provide an elegant geometric interpretation of the inhomogeneity associated with the Weyl part of curvature tensors and scalars in terms of fluctuations of the Ricci scalar. 

Since a non--vanishing shear and Weyl curvature are indicative of the difference between LTB and FLRW models, it is very useful to express the basic local perturbations $\Drho$ and $\Dh$ as     
\bse\ba \Drho =\frac{\rho-\rho_q}{\rho_q}=-\frac{\EE}{ 4\pi\rho_q/3}= \frac{6\Psi_2}{\CR-6\Psi_2}=\frac{\xi}{1-\xi},\label{drho2}\\
\Dh = \frac{\HH-\HH_q}{\HH_q}=-\frac{\Sigma}{\HH_q}=-\frac{\Sigma}{\HH+\Sigma}=-\frac{\zeta}{1+\zeta},\label{dh2}\ea\ese
where the quotients
\begin{equation} \xi \equiv \frac{6\Psi_2}{\CR},\qquad \zeta \equiv \frac{\Sigma}{\HH},\label{xizeta} \end{equation}
provide an invariant measure of the deviation from FLRW geometry through the ratio of Weyl to Ricci scalar curvatures ($\xi$) and the ratio of anisotropic to isotropic expansion ($\zeta$)
\footnote{Wainwight and Andrews \cite{wainwright} also emphasize the role of the ratio of Weyl to Ricci curvature as an invariant measure of the deviation of LTB models from FLRW geometry. However, their study was not based on averaging and they consider a quadratic non--negative ratio $C_{abcd}C^{abcd}/\CR_{ab}\CR^{ab}$. As a contrast, the sign of the first order ratio $\Psi_2/\CR$ that we use here is not {\it a priori} defined. We would like to emphasize that nonzero $\xi$ and $\zeta$ in (\ref{xizeta}) do not provide a measure of ``inhomogeneity'' in general, as both are nonzero in homogeneous but anisotropic Bianchi models and $\xi=0$ holds for conformally flat spacetimes, which are (in general) inhomogeneous \cite{kras1,stephani}.}
. The remaining perturbations $\DKK$ and $\DOm$ can be expressed in terms of $\xi$ and $\zeta$ by using the constraints (\ref{Dhdef}) and (\ref{DOmdef}) to eliminate them in terms of $\Drho$ and $\Dh$ and then substituting (\ref{drho2}) and (\ref{dh2}). 

Since $\Sigma$ and $\EE$ are local linear fluctuations of $\HH$ and $\rho$, we have for all $r=r_b$:
\begin{equation}  \langle \Sigma \rangle_q[r_b]\ne 0,\qquad \langle\EE \rangle_q[r_b]  \ne 0, \end{equation}
However, as a consequence of (\ref{var}), (\ref{cov}) and (\ref{SigE2}), the averages of the scalars $\Sigma^2,\,\EE^2$ and $\Sigma \,\EE$ are averages of local quadratic fluctuations, leading to the following statistical fluctuations that relate to the variance moment with respect to $\HHav_q$ and $\rhoav_q$ at any domain: 
\bse\ba \fl \langle\Sigma^2\rangle_q =\langle\, \left[\Del(H)\right]^2\,\rangle_q = \langle \HH^2\rangle_q - \HHav_q^2 = \hbox{{\bf Var}}_q(\HH),\label{varHH}\\
\fl \langle\EE^2\rangle_q =\left[\frac{4\pi}{3}\right]^2\langle\,\left[\Del(\rho)\right]^2\,\rangle_q = \left[\frac{4\pi}{3}\right]^2\left[\langle \rho^2\rangle_q - \rhoav_q^2\right] = \left[\frac{4\pi}{3}\right]^2\hbox{{\bf Var}}_q(\rho),\label{varrho}\\
\fl \langle \EE\,\Sigma\rangle_q = \frac{4\pi}{3}\langle \Del(\rho)\,\Del(\HH)\rangle_q =\frac{4\pi}{3}\left[\,\langle \rho\,\HH\rangle_q-\rhoav_q\,\HHav_q\right]= \frac{4\pi}{3}\,\hbox{{\bf Cov}}_q(\rho,\HH),\label{covrhoHH}
\ea\ese     
so that, in view of (\ref{Sigma1}) and (\ref{contr3a}), these fluctuations relate to averages of quadratic contractions of the shear and electric Weyl tensors:
\bse\ba \langle \sigma_{ab}\sigma^{ab}\rangle_q = 6\langle\Sigma^2\rangle_q = 6\hbox{{\bf Var}}_q(\HH),\label{varsigsq}\\
\langle E_{ab} E^{ab}\rangle_q = 6\langle\EE^2\rangle_q = 6\langle(\Psi_2)^2\rangle_q = \frac{32\pi^2}{3}\hbox{{\bf Var}}_q(\rho), \label{varEEsq}\\
\langle \sigma_{ab} E^{ab}\rangle_q = 6\langle \Sigma\,\EE\rangle_q = 8\pi \hbox{{\bf Cov}}_q(\rho,\HH),\label{covsigEE}\ea\ese
while the q--averages of contractions of the Riemann and Weyl tensors are expressible either in terms of $\rho$ or $\CR$ and their statistical variance moments:
\bse\ba 
\fl\langle\CR_{abcd}\CR^{abcd}\rangle_q=\frac{256\pi^2}{3}\left[\hbox{{\bf Var}}_q(\rho)+\frac{5}{4}\langle\rho^2\rangle_q\right] =\frac{4}{3}\left[\hbox{{\bf Var}}_q(\CR)+\frac{5}{4}\langle \CR^2\rangle_q\right],\label{Kretsch}\\
\fl \langle C_{abcd}C^{abcd}\rangle_q = \frac{256\pi^2}{3}\,\hbox{{\bf Var}}_q(\rho)= \frac{4}{3}\,\hbox{{\bf Var}}_q(\CR) = 8\langle E_{ab}E^{ab}\rangle_q,\label{sqWeyl}
\ea\ese
where we used the fact that $\langle{\CR}\rangle_q =8\pi\rhoav_q$ and $\CR_{ab}\CR^{ab}=\CR^2$. We can express the quadratic ratio of Weyl to Ricci curvatures as a sort of ``standard deviation'' of $\rho$ with respect to $\rhoav_q$:
\begin{equation}\fl  \frac{6\langle E_{ab}E^{ab}\rangle_q}{\langle \CR_{ab}\CR^{ab}\rangle_q}=\frac{6\langle(\Psi_2)^2\rangle_q}{\langle(\CR)^2\rangle_q}=\frac{6\hbox{{\bf Var}}_q(\rho)}{\langle\rho^2\rangle_q} = \frac{\langle \rho^2\rangle_q-\rhoav_q^2}{\langle\rho^2\rangle_q},\end{equation}
A similar standard deviation of $\HH$ with respect to $\HHav_q$ follows as the quotient of averages of quadratic covariant scalars $\sigma_{ab}\sigma^{ab}$ and $\HH^2=\theta^2/9$:
\begin{equation}\fl  \frac{\langle \sigma_{ab}\sigma^{ab}\rangle_q}{6\langle \HH^2\rangle_q}=\frac{\hbox{{\bf Var}}_q(\HH)}{\langle\HH^2\rangle_q} = \frac{\langle \HH^2\rangle_q-\HHav_q^2}{\langle\HH^2\rangle_q},\end{equation}
where we used (\ref{Sigma1}) and (\ref{SigE2}). 

\section{Comparison with Buchert's averaging.}

\subsection{Scale factors and reference volumes.}

The q--average of the Hubble factor in equations (14b) and (\ref{rhoHKq}) can also be written as the definition of a dimensionless scale factor associated with (\ref{qave}):
\begin{equation} \frac{\dot a}{a}=\HH_q,\qquad a\equiv \frac{R}{R_0},\label{aavq}\end{equation}
where $R_0=R(t_0, r)$ and $t_0$ is a fixed arbitrary value of $t$. A scale factor analogous to $a$ (denoted as ``$a_{\DD}$'') is also defined in Buchert's averaging formalism for every domain \cite{buchert,buchGRG,buchCQG}. In its application to LTB models \cite{LTBave1,LTBave2,mattsson,sussBR,sussIU,suss2011} the Buchert's scale factor for a domain $\DD[r_b]$ takes a different form from (\ref{aavq}): 
\begin{equation} a_{\DD} = \left[\frac{\VV_p}{\VV_{p0}}\right]^{1/3},\qquad \VV_p=\int_{\DD}{\dd \VV_p}=4\pi\int_0^r{\frac{R^2 R'\dd\bar r}{\FF}},\label{aavp}\end{equation}
where $\dd \VV_p$ is the proper volume element (\ref{dV}), $\FF$ is the weight factor (\ref{FF}) and $\VV_{p0}$ is the proper volume of any given domain $\DD[r]$ evaluated at an arbitrary $t=t_0$. Comparing (\ref{aavq}) and (\ref{aavp}) leads to the relation: $a =[\VV_q/\VV_{q0}]^{1/3}$, where $\VV_q\equiv \int_{\DD[r]}{\FF\dd\VV_p}=(4\pi/3)R^3$ can be regarded as a ``quasi--local volume'' of the domain $\DD[r]$. The quasi--local volume follows from an integral carried on at arbitrary time slices ($t$ fixed), therefore, $R=\int{\dd R}=\int{ R'\dd \bar r}$ can always be used as a global radial coordinate along the time slices in models in which these slices admit a symmetry center (at $r=0$) and have ``open'' topology, so that $R'>0$ holds for all $r$ (assuming absence of shell crossings). As a consequence, $\VV_q$ is equivalent to an Euclidean 3--volume integral at each time slice for these models. This is not the case for models in which the time slices admit two symmetry centers and have spherical topology (homeomorphic to $\mathbb{S}^3$) because $R'$ changes sign regularly and $R$ cannot be a global radial coordinate for these slices. In this case $\VV_q$ reaches a maximal value at the ``equator'' of the slice and vanishes at the second symmetry center, so that we have $\VV_q=0$ for the whole slice. Evidently, a non--trivial $\FF$ in (\ref{aavp}) implies that the reference volume of Buchert's average $\VV_p$ is not equivalent to an Euclidean 3--volume integral and $\VV_p> 0$ if evaluated for any whole slice with spherical topology.

\subsection{No back--reaction in the q--average.}   

The most important difference between the q--average (\ref{qave}) and Buchert's averaging formalism  is the fact that the evolution equations that follow from (\ref{qave}) and its fluctuations lack the ``back--reaction'' correlation terms that appear in Buchert's  evolution equations \cite{buchert,buchGRG,buchCQG}. This fact follows readily by comparing (\ref{FLRW1})--(\ref{FLRW2}) with the equivalent equations in Buchert's formalism as applied to LTB models \cite{LTBave1,LTBave2,mattsson,sussBR,sussIU,suss2011}. 

Using (\ref{SigE2}) to express $\Sigma$ in terms of $\HH-\HH_q=\Del(\HH)$ and the commutation rule $\langle \HH\rangle\dot{}_q-\langle\dot \HH\rangle_q$ taken from  (\ref{tder}), the weighed average (\ref{qave}) applied to the Raychaudhuri equation (\ref{EV1}) yields: 
\begin{equation} \fl \langle \HH\rangle\dot{}_q[r_b] =-\langle \HH\rangle_q^2[r_b]-\frac{4\pi}{3}\langle\rho\rangle_q[r_b]+2\QQ_q[r_b],\label{raychq}\end{equation}
where $\QQ_q[r_b]$ is the back--reaction term:
\ba
\fl \QQ_q[r_b] \equiv \langle\left(\HH-\HHav_q\right)^2\rangle_q-\langle\left(\HH-\HH_q\right)^2\rangle_q =
\langle \left[\DENL(\HH)\,\right]^2\rangle_q[r_b] - \langle \left[\Del(\HH)\right]^2\rangle_q[r_b]=0,\nonumber\\\label{zeroQ}\ea
which vanishes identically for every domain as a consequence of (\ref{var}) applied to $A=\HH$ (see (\ref{B11})--(\ref{B13}) for the proof of this result, see also \cite{suss2011}). Hence, (\ref{raychq}) reduces exactly to the FLRW Raychaudhuri equation in (\ref{FLRW2}) with zero back--reaction.  

This is a completely different outcome in comparison with the average from Buchert's formalism, which we will denote henceforth with the subindex ${}_p$ as $\langle A\rangle_p[r_b]$. Applying Buchert's average to the Raychaudhuri equation (\ref{EV1}) yields a similar equation as (\ref{raychq}), but with a non--vanishing back--reaction term \cite{buchert,buchGRG,buchCQG}
\ba \fl\langle \HH\rangle\dot{}_p =-\HHav_p^2-\frac{4\pi}{3}\rhoav_p+2\QQ_p[r_b],\nonumber\\ 
\hbox{with}:\qquad \QQ_p[r_b]\equiv \langle\left(\HH-\HHav_p\right)^2\rangle_p-\langle\left(\HH-\HH_q\right)^2\rangle_p\ne 0,
\label{nonzeroBR}\ea
with the Hamiltonian constraint (\ref{aveFriedman}) taking the form
\begin{equation}\HHav_p^2[r_b] = \frac{8\pi}{3}\rhoav_p[r_b] -\KKav_p[r_b]-\QQ_p[r_b].\end{equation}
and the energy balance equation (\ref{EV2}) remaining with the same form.    

Another important difference with Buchert's averaging is the issue of completeness of the evolution equations. If we consider averaged scalars $\Aav_p$ and fluctuations constructed with this average in the framework of Buchert's formalism, we do not obtain a complete self--consistent set of evolution equations unless we make extra assumptions on the back--reaction terms \cite{buchGRG,buchCQG,buchdark}. As a contrast, the scalars that emerge from the q--average (as functionals and as functions) yield complete self--consistent sets of evolution equations without the need to introduce further assumptions (these evolution equations are derived in section 4 of part II).

Evidently, since invariant scalars like $\sigma_{ab}\sigma^{ab},\, E_{ab}E^{ab}$ and $\sigma_{ab}E^{ab}$ are quadratic fluctuations with respect to the q--functions $\rho_q,\,\HH_q$ (which relate to the q--averages $\rhoav_q$ and $\HHav_q$), and not with respect to functions associated with $\rhoav_p$ and $\HHav_p$,  quadratic and higher order fluctuations in Buchert's formalism are not equal to these invariants, and thus we cannot obtain in this formalism the simple straightforward relations between averages of these invariants and statistical moments of $\rho$ and $\HH$ of section 6. However, we can still obtain analytic expressions for computing Buchert's average for these invariants. Considering that for any domain the q--average and Buchert's average of any scalar are related by
\begin{equation} \Aav_q[r_b] = \frac{\langle A\,\FF\rangle_p[r_b]}{\langle \FF\rangle_p[r_b]}, \end{equation} 
together with the identities (\ref{varHH})--(\ref{covrhoHH}) and (\ref{varsigsq})--(\ref{covsigEE}), we can obtain the following exact expressions for Buchert's average of quadratic scalar contractions:
\bse\ba 
\fl\langle \sigma_{ab}\sigma^{ab}\rangle_p = \frac{\langle\HH^2\,\FF\rangle_p}{\langle \FF\rangle_p}-\frac{\langle\HH\,\FF\rangle^2_p}{\langle \FF\rangle^2_p}-\frac{\hbox{{\bf Cov}}_p(\sigma_{ab}\sigma^{ab},\,\FF)}{\langle \FF\rangle_p},\label{sigsigp1}\\
\fl \langle E_{ab} E^{ab}\rangle_p = \frac{32\pi^2}{3}\left[\frac{\langle\rho^2\,\FF\rangle_p}{\langle \FF\rangle_p}-\frac{\langle\rho\,\FF\rangle^2_p}{\langle \FF\rangle^2_p}\right]-\frac{\hbox{{\bf Cov}}_p(E_{ab}E^{ab},\,\FF)}{\langle \FF\rangle_p},\label{EEp1}\\
\fl \langle \sigma_{ab} E^{ab}\rangle_p = 8\pi\left[\frac{\langle\rho\,\HH\,\FF\rangle_p}{\langle \FF\rangle_p}-\frac{\langle\rho\,\FF\rangle_p\langle\HH\,\FF\rangle_p}{\langle \FF\rangle^2_p}\right]-\frac{\hbox{{\bf Cov}}_p(\sigma_{ab}E^{ab},\,\FF)}{\langle \FF\rangle_p},\label{Esigp1}
\ea\ese
which lead to the following appealing identities which relate the two averages of these scalars and their correlation with the weight factor $\FF$:
\bse\ba 
\langle \sigma_{ab}\sigma^{ab}\rangle_q - \langle\sigma_{ab}\sigma^{ab}\rangle_p = \frac{{\bf {Cov}}_p(\sigma_{ab}\sigma^{ab},\,\FF)}{\langle \FF\rangle_p},\label{sigsigp2}\\
\langle E_{ab} E^{ab}\rangle_q - \langle E_{ab}E^{ab}\rangle_p = \frac{{\bf {Cov}}_p(E_{ab}E^{ab},\,\FF)}{\langle \FF\rangle_p},\label{EEp2}\\
\langle \sigma_{ab} E^{ab}\rangle_q - \langle \sigma_{ab}E^{ab}\rangle_p = \frac{{\bf {Cov}}_p(\sigma_{ab}E^{ab},\,\FF)}{\langle \FF\rangle_p},\label{Esigp2}
\ea\ese
where we have omitted the $[r_b]$ symbol to simplify notation and ${\bf {Cov}}_p$ is the covariance correlation moment (\ref{cov}) with respect to Buchert's average. 

\section{Gravitational entropy.}

A gravitational entropy functional has been proposed by Hosoya, Buchert and Morita   \cite{buchGRG,entropy1,entropy2,entropy3}, as an application  of the Kullback--Leibler functional of Information Theory to spacetimes with inhomogeneous dust sources. While Hosoya {\it et al.}  defined this functional with Buchert's average ({\it i.e.} (\ref{qave}) with $\FF=1$ acting on any scalar), we consider here the same functional with the weighed q--average (\ref{qave}) with $\FF$ not equal to a constant. Hence, our approach will yield different theoretical connections.

Following \cite{buchGRG,entropy1,entropy2,entropy3}, we can define the Kullback--Leibler entropy functional in terms of the q--average over a domain $\DD[r_b]$ as
\ba\fl S-\Sequ=\gamma_0\int_{\DD[r_b]}{\rho\, \ln \left[\frac{\rho}{\rhoav_q[r_b]}\right]\FF\dd\VV_p}=\gamma_0\left\langle \rho \ln \frac{\rho}{\rhoav_q[r_b]}\right\rangle_q[r_b]\,\VV_q(r_b),\label{Sdef}\\
\fl \hbox{with}:\qquad \VV_q = \int_{\DD[r_b]}{\FF\,\dd\VV_p}=\frac{4\pi}{3}R^3(r_b),\label{qvol}\ea
where $\Sequ=\Sequ( r_b)$, and we have introduced the constant $\gamma_0=k_{_B}/(mc^2)$, with $k_{_B}$ being the Boltzmann's and $m$ a particle mass, so that $S-\Sequ$ has units of entropy. If we define the entropy current associated with (\ref{Sdef}) as $S^a = \rho\,(S-\Sequ)\,u^a$, then by considering the rest mass density conservation law (\ref{EV2}) we obtain the following entropy balance law:
\begin{equation} \nabla_a S^a \geq 0 \qquad \Rightarrow\qquad \dot S \geq 0,\label{Sbal}\end{equation}
which allows us to identify the inhomogeneity measured by $\rho( r)\ne \rhoav_q[r_b]$ as the source of gravitational entropy production ($\dot S>0$) in arbitrary domains $\DD[r_b]$ of LTB models, with equilibrium states complying with $\dot S=0,\,S=\Sequ$ and corresponding to:
\begin{description}
\item FLRW models as the homogeneous subset (particular case) of LTB models for which $\rho( r)=\rhoav_q[r_b]$ holds for all $r,\,r_b,\,t$. 
\item Specific boundaries of generic LTB models, such as:
\begin{itemize}
\item The center worldline (in models admitting a regular center) corresponding to the case $r_b=0$, since $\rho(t,0)=\rho_q(t,0)$ holds for all $t$ \cite{RadAs,RadProfs,suss2011}.
\item Asymptotic boundary $r\to\infty$ along radial rays of time slices $\T[t]$ of LTB models radially converging to a FLRW state (see section 10.2 and \cite{RadAs}), for which both $\rho$ and $\rho_q$ tend to the same value $\rho_{_\infty}=\tilde\rho(t)$ in this limit (see figure 4 of part II).
\item The boundary $r=r_b$ (for all $t$) of Swiss cheese holes in domains $\DD[r_b]$ smoothly matched with a FLRW region (see sections 4.4 and 10.1, and panels (b) and (d) of figures 1, 2 and 3 of part II). Evidently $\dot S=0$ holds for all $t$ in the FLRW region $r>r_b$.
\end{itemize}   
\end{description} 
We remark that in all the  equilibrium states listed above we have $\Del(\rho)= \DENL(\rho)=0$, and thus entropy production is closely related with the existence of nonzero density fluctuations.  For the remaining of this section we will assume domains such that $\dot S\ne 0$ holds.  

In order to evaluate $\dot S$, we apply the time derivative commutation rule (\ref{tder}) to (\ref{Sdef}), leading after some algebraic manipulation to the same relation between $\dot S$ and the non--commutativity of the time derivative and the average found in \cite{entropy1,entropy2,entropy3} for Buchert's average:
\begin{equation}\frac{\dot S(r_b)}{\gamma_0\VV_q(r_b)}=\langle \dot\rho\rangle_q[r_b]-\rhoav\dot{}_q[r_b], \label{Sdot1}\end{equation}
which, with the help of (\ref{tder}) and (\ref{cov}), can be related to the negative statistical correlation  of fluctuations of $\rho$ and $\HH$, given by the covariance momentum with respect to the q--averages of these variables in a domain $\DD[r_b]$: 
\begin{equation}\fl \frac{\dot S}{\VV_q}=-3\gamma_0\left[\,\langle\rho\HH\rangle_q-\rhoav_q\HHav_q\,\right]=-3\gamma_0\hbox{{\bf Cov}}_q(\rho,\HH)=-3\gamma_0\langle \Del(\rho)\,\Del(\HH)\rangle_q\geq 0, \label{Sdot2}\end{equation}
so that:
\begin{equation}\hbox{{\bf Cov}}_q(\rho,\HH)= \langle\Del(\HH)\Del(\rho)\rangle_q[r_b]<0\,\,\Rightarrow\,\, \dot S(r_b)>0,\label{condDhDr}\end{equation}
where we have applied (\ref{cov}) to replace non--local fluctuations with local ones (see proof in Appendix C). Considering (\ref{Sigma1})--(\ref{EE1}), (\ref{SigE2}),  (\ref{contr3a}) and (\ref{covrhoHH}), condition (\ref{condDhDr}) can be given in terms of the q--average of a scalar invariant by:
\begin{equation} \langle \sigma_{ab}\,E^{ab}\rangle_q[r_b]<0\,\,\Rightarrow\,\, \dot S(r_b)>0,\label{geom}\end{equation}
which is a very elegant way to connect (\ref{Sdef}) with an unequivocal and completely coordinate independent  marker of inhomogeneity, as it contains contributions from density and velocity fluctuations.  Notice that, as consequence of (\ref{Sdot2})--(\ref{geom}), the condition $\dot S=0$ in equilibrium states implies $\Del(\rho)=\DENL(\rho)=\Del(\HH)=\DENL(\HH_q)=0$, and thus fluctuations of all q--scalars necessarily vanish in these states. Also, these equilibrium states imply that the local quadratic fluctuation and variance moment of the Ricci scalar is zero (from (\ref{ECricci}) and (\ref{sqWeyl})).       

As pointed out by the authors in references \cite{entropy1,entropy2,entropy3} (who also obtained the relations (\ref{Sdot2}) and (\ref{condDhDr}) but not (\ref{geom})), a negative correlation of the density and Hubble fluctuations seems to be consistent with the intuitive behavior of gravitational clustering processes in structure formation scenarios: over/under dense regions tend to contract/expand as density increases/decreases.   However, this intuitive behavior occurs in the time evolution, not (necessarily) in domains along time slices where averages are computed. Therefore, as much as (\ref{condDhDr}) is an intuitively appealing and elegant proposition (in view of (\ref{geom})), it is not {\it a priori} evident that we will obtain the desired negative sign of $\Del(\rho)\Del(\HH)$ when actually computing the fluctuations. Since verifying (\ref{condDhDr}) in arbitrary domains of generic inhomogeneous LTB models is a comprehensive task that is beyond the scope of this paper, we provide in this section a guideline on how this verification can be accomplished, as well as some preliminary results.

Testing (\ref{condDhDr}) involves computing radial integrals of quantities whose explicit radial dependence is not known (and thus must be done numerically). However, we can infer the fulfillment of (\ref{condDhDr}) by means of the following sufficient (but not necessary) condition that can be evaluated at the boundary of every domain $r=r_0$:
\begin{equation}  \fl \Del(\HH)\Del(\rho) = [\,\HH(r_b)-\HH_q(r_b)\,][\,\rho(r_b)-\rho_q(r_b)\,]<0\,\,\Rightarrow\,\, \dot S(r_b)>0,\label{sufcond}\end{equation} 
where we used (\ref{cov}) (see its proof in (\ref{B31})--(\ref{B32})) to be able to consider the fluctuations evaluated at the boundary $r=r_b$ of a fixed but arbitrary domain. Considering (\ref{rder}) and (\ref{intparts}), the product of these fluctuations is 
\begin{equation} \fl \Del(\HH)\Del(\rho)=\frac{\rho_q'(r_b)\,\HH_q'(r_b)}{[\VV'_q(r_b)/\VV'_q(r_b)]^2} = \frac{1}{\VV_q^2(r_b)}\int_0^{r_b}{\rho'\,\VV_q\,\dd\bar r}\,\int_0^{r_b}{\HH'\,\VV_q\,\dd\bar r},\label{DHDr}\end{equation}
and thus condition (\ref{sufcond}) can be given directly in terms of the correlation of the radial gradients of the density and the Hubble expansion scalar:
\footnote{The condition $\rho'( r)\,\HH'( r)< 0$ is also a sufficient condition if it holds for the full integration range $0<r<r_b$ in the integrals in (\ref{DHDr}). On the other hand, conditions (\ref{dotSgrad1})  and (\ref{dotSgrad2}) only need to be evaluated at $r_b$. Notice that for monotonic profiles $\rho'(r_b)<0$ and $\rho'(r_b)>0$ respectively imply $\rho'_q(r_b)<0$ and $\rho'_q(r_b)>0$, but the converses are false, see \cite{RadProfs}. }
\begin{equation} \rho_q'(r_b)\,\HH_q'(r_b)< 0\qquad\Rightarrow\qquad \dot S(r_b)>0,\label{dotSgrad1}\end{equation}
or (from (\ref{Dagrad})) in terms of the local perturbations $\Drho,\,\Dh$:
\begin{equation} \Drho(r_b)\,\Dh(r_b) < 0 \qquad\Rightarrow\qquad \dot S(r_b)>0.\label{dotSgrad2}\end{equation}
Either one of (\ref{dotSgrad1}) or (\ref{dotSgrad2}) allows us to test (at least qualitatively) the fulfillment of $\dot S> 0$ in arbitrary domains of generic LTB models by means of the analytic and qualitative results derived in \cite{suss2011,RadAs,RadProfs}. While a detailed verification of (\ref{dotSgrad1}) or (\ref{dotSgrad2}) for generic LTB models is beyond the scope of this paper, we provide below a list the preliminary results: 
\begin{description}
\item[Entropy production is positive.] Conditions (\ref{dotSgrad1}) and (\ref{dotSgrad2}) hold in the following cases: 
\begin{itemize}
\item Late time evolution of elliptic models (see (\ref{ellt1})) in time slices that ``hit''  the collapsing singularity. This result can be justified by qualitative arguments (see figure 1).
\item Asymptotic time evolution ($t\to\infty$ for all $r$) of hyperbolic models (see (\ref{ellt1})). This result follows from the fact that $\Dh \to \DKK/2$ holds in this limit for these models (because $\Omega_q\to 0$), and thus from table 1 of \cite{RadProfs} we have $\Drho\Dh<0$ for all initial conditions complying with regularity. 
\end{itemize}
\item[Entropy production is negative.] Conditions (\ref{dotSgrad1}) and (\ref{dotSgrad2}) are violated in the following cases:
\begin{itemize}
\item Parabolic models (spatially flat models $\KK=0,\,\Omega_q=1$, see first class of solutions in (\ref{ellt1})). Since $\Dh=\Drho/2$ holds for all $t$, and regularity conditions require $-1<\Dh\leq 0$ (see \cite{RadAs,RadProfs}), then $\Drho\Dh>0$ holds for all $t$.
\item Early time evolution of elliptic and hyperbolic models with a non--simultaneous big bang. This result can be justified by qualitative arguments (see figure 1).   
\end{itemize}
 \end{description}
These results clearly indicate an important connection between the failure to fulfill a positive entropy production and specific conditions in which the density decaying modes dominate over the growing modes (parabolic models and near a non--simultaneous big bang \cite{kras1,kras2,BKHC2009,zibin,wainwright}). This is an important theoretical feature worth a comprehensive examination to be undertaken in future work.
\begin{figure}
\begin{center}
\includegraphics[scale=0.40]{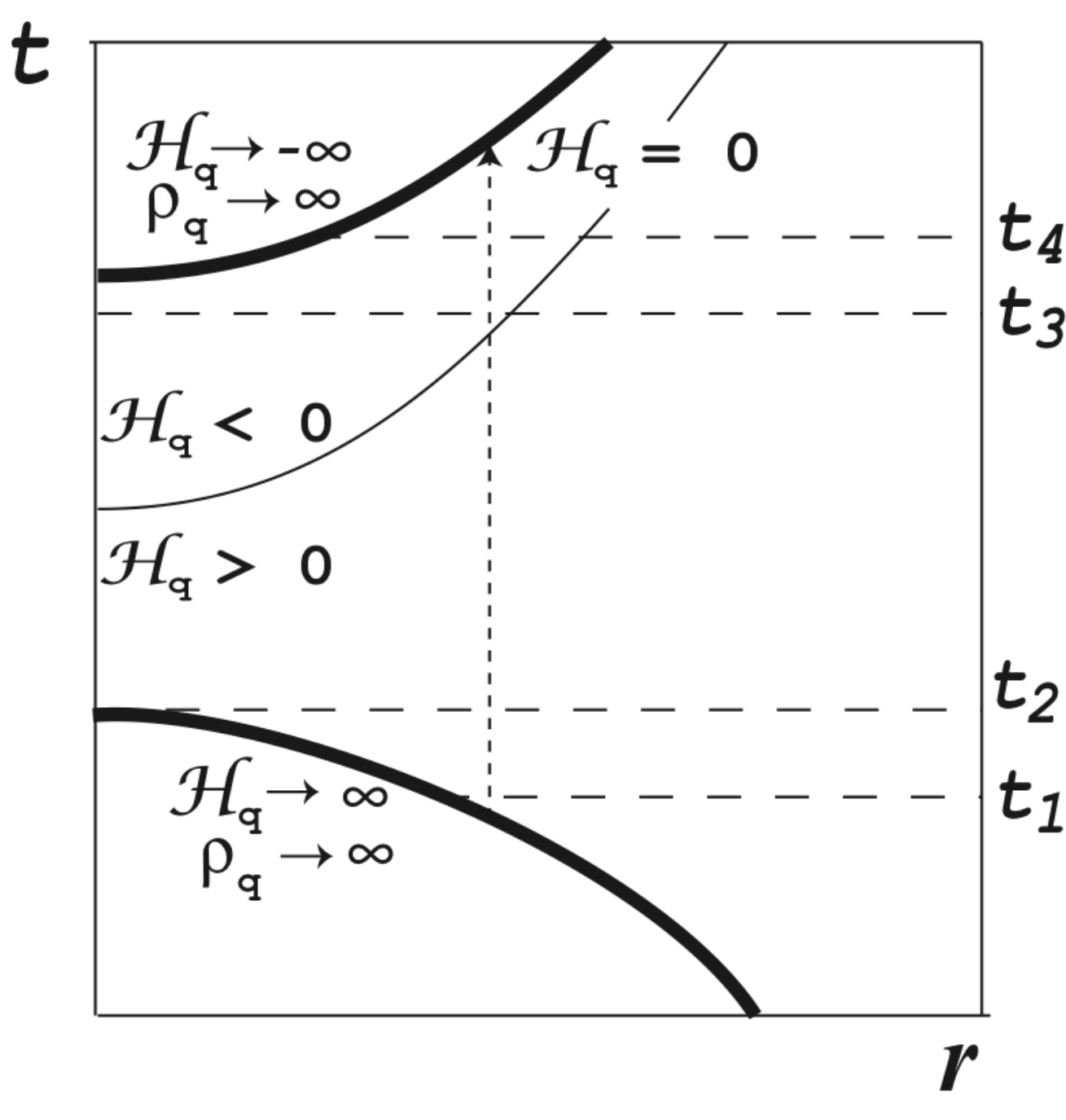}
\caption{{\bf Time slices and the expanding and collapsing singularities.} The figure displays time slices $t_1,\,t_2,\, t_3,\,t_4$ (long dashed horizontal lines)  in the $(t,r)$ plane and the worldline of a comoving layer $r=$ constant (short dashed vertical line) evolving from the a non--simultaneous big bang (thick lower curve $\tbb( r)$) towards the collapsing singularity (thick upper curve $\tcoll( r)$) in an elliptic model. For $t_1,\,t_2$ we have necessarily $\HH'_q <0$ and $\rho'_q<0$, whereas we have $\HH'_q >0$ and $\rho'_q<0$ for $t_3,\,t_4$. Therefore, conditions (\ref{dotSgrad1}) and (\ref{dotSgrad2}) hold for slices near $\tcoll$ but fails for slices near $\tbb$. The latter failure also occurs in hyperbolic models with non--simultaneous big bang that lack a collapsing singularity. }
\label{fig6}
\end{center}
\end{figure}

Besides the balance law (\ref{Sbal}), it is useful to compute time concavity/convexity of $S$ given by the sign of $\ddot S$, which follows readily by applying (\ref{tder}) to (\ref{Sdot2}):
\ba\frac{\ddot S}{\VV_q} &=& \frac{9\gamma_0}{2}\left\langle \rho[\Del(\HH)]^2+\frac{2}{3}\HH_q\Del(\rho)\Del(\HH)+\frac{4\pi}{9}[\Del(\rho)]^2\right\rangle_q\nonumber\\
&=& \frac{3\gamma_0}{4}\left\langle \rho\,\sigma_{ab}\sigma^{ab}+\frac{1}{2\pi}\HH_q\sigma_{ab}E^{ab}+\frac{1}{4\pi}E_{ab}E^{ab}\right\rangle_q,\label{ddotS}\ea
where we used (\ref{EV1})--(\ref{EV2}), (\ref{FLRW1})--(\ref{FLRW2}) and (\ref{SigE2})  to eliminate the derivatives $\dot\rho,\,\dot\rho_q,\,\dot\HH,\,\dot\HH_q$ and $\Sigma$ in terms of fluctuations $\Del(\rho),\,\Del(\HH)$.  

\section{Conclusion and summary.}

We have considered the q--average (the scalar average defined by (\ref{qave})) in spherical comoving domains $\DD[r]$ for the study of generic LTB models and their properties. This average generalizes the standard proper volume average of Buchert's formalism by the presence of a non--trivial weight factor ($\FF$) in the definition (\ref{qave}) (the q--average and Buchert's average of a given scalar $A$ only coincide if $\FF=$ const. \cite{suss2011}). 

Evidently, the q--average with the specific weight factor (\ref{FF}) is specially suited for LTB models (though it also works for LTB models with pressure \cite{sussQL,suss2009} and Szekeres models \cite{sussbol,peel}), while Buchert's average is readily applicable to any spacetime and any foliation in time slices. In this respect, the q--average is ``model dependent'' and more restrictive than Buchert's averaging formalism. However, it still remains to verify if useful results emerge from a weighed scalar average that may be devised for more general (or even fully general) spacetimes.

By applying the q--average to LTB covariant scalars $A$, we obtained  new dynamical variables (q--scalars) that can be defined as functionals $\Aav_q[r_b]$ (the average is assigned to a whole fixed but arbitrary domain) or as functions $A_q$ (q--functions that depend on the average for domains with varying boundaries). The q--scalars are covariant, as they are expressible in terms of curvature and kinematic invariants (see section 6), and also have very special and useful properties. Specifically, the q--scalars that are common with FLRW models ($A_q$ and $\Aav_q[r_b]$ for $A=\rho,\,\HH,\,\KK,\,\Omega$) identically satisfy FLRW fluid flow dynamics and (as we show in part II) their analytic forms mimic FLRW expressions that are commonly used in applications of LTB models (specially in void models).  The q--scalars give rise to fluctuations and perturbations that are local ($\Del(A),\,\Da$) or non--local ($\DENL(A),\,\Daav$), depending on whether they are, respectively, defined for q--scalars that are functions or functionals.  The following results are worth highlighting:
\begin{itemize}
\item All proper tensors (curvature and kinematic) characteristic of LTB models are expressible by irreducible algebraic expansions of $g_{ab},\, u^b$ and the common eigenframe of the shear and electric Weyl tensors ($\hbox{\bf{e}}^a_b$), with coefficients given by the q--scalars $\rho_q,\,\HH_q,$ and their relative local fluctuations $\Del(\rho),\,\Del(\HH)$ (these q--scalars fluctuations can be expressed in terms of the q--scalars $\KK_q,\,\Omega_q$ and their fluctuations through the constraints (\ref{qFriedman}), (\ref{Omdef}) and (\ref{Dhdef})--(\ref{DOmdef})). In particular, all curvature tensors are expressible in terms of these algebraic expansions with coefficients given by the Ricci scalar $\CR$ and the conformal invariant $\Psi_2$ (which is its local fluctuation $\Del(\CR)$) (section 6).
\item Quadratic local fluctuations of $\rho$ and $\HH$ are equal to invariants that follow from contractions of the tensors involved in the dynamics of LTB models: the Ricci, electric Weyl and shear tensors. The q--averages of these fluctuations (local and non--local) are equal to statistical moments (variance and correlation moments) of $\rho$ and $\HH$ (see section 6). 
\item Local perturbations $\Da$ provide an elegant and invariant characterization of the deviation from FLRW geometry  in terms of the ratio of Weyl to Ricci curvature: $\Psi_2/\CR$, where $\Psi_2$ is the only nonzero Newman--Penrose conformal invariant and $\CR$ is the Ricci scalar, whereas relative Hubble scalar  fluctuations $\Dh$ do so in terms of the ratio of anisotropic vs. isotropic expansion: $\Sigma/\HH$, where $\Sigma$ is the eigenvalue of the shear tensor. All q--scalars can be given in terms of these scalar invariants (section 6). 
\item The basic difference with the standard proper volume average of Buchert's formalism is the absence of back--reaction correlation terms in the evolution equations associated with q--averages and their perturbations. Also, the relation between the averages of quadratic and higher order fluctuations and averages of scalar invariants and statistical moments of $\rho$ and $\HH$ does not occur with Buchert's average. We derive exact expressions for the Buchert's average of quadratic scalar invariants (see section 7).  

\end{itemize}

As a quick application, we examined in section 8 the definition of gravitational entropy based on the Leibler--Kullback functional from Information Theory, which has been proposed by Hosoya, Buchert and Murita  \cite{buchGRG,entropy1,entropy2,entropy3} using Buchert's averages. Considering this functional as given in terms of q--averages yields the same conjecture obtained in these references, namely: a relation should exist between positive entropy production ($\dot S>0$) and a negative correlation between fluctuations of $\rho$ and $\HH$. However, testing this conjecture is easier if we use q--averages, since: $\dot S$ is proportional to the q--average of the scalar invariant $\sigma_{ab}E^{ab}$, and also $\rhoav_q$ and $\HHav_q$ are expressible in closed analytic forms and the evolution equations for their fluctuations and perturbations (see part II) do not involve complicated back--reaction terms. Hence, we were able to obtain the sufficient conditions for $\dot S>0$ in terms of closed analytic expressions, such as (\ref{dotSgrad1}) and (\ref{dotSgrad2}), that can be tested in arbitrary domains of generic LTB models.      

As a comparison with our approach and preliminary results, a recent perturbative study \cite{entropy3} of the Leibler--Kullback functional in the context of Buchert's average yields $\dot S$ only in terms of the average (Buchert's average) of the invariant scalar $C_{abcd}C^{abcd}$ (which is $8E_{ab}E^{ab}$ in LTB models), hence it excludes the contribution to inhomogeneity from the shear tensor (a spatial gradient of the velocity in the conformal Newtonian gauge) that comes from the scalar invariant $\sigma_{ab}E^{ab}$. Hosoya {\it et al} have not yet applied the Leibler--Kullback functional to generic LTB models in a non--perturbative manner (only a very simple example of an LTB model was considered in \cite{entropy2}, but this model exhibits shell crossings). Although  they obtained conditions that are analogous to (\ref{condDhDr}), their main interest has been to provide a theoretical connection between the growth of structure complexity associated with $\dot S>0$ and the dynamical implications of the back--reaction terms that appear in Buchert's formalism. Since the back--reaction terms of Buchert's formalism vanish for q--averages, our results provide a different theoretical perspective, namely: the connection between $\dot S>0$ in (\ref{Sdot2}) and $\ddot S$ in (\ref{ddotS}) associated with the growth of structure complexity and the statistical moments that arise from the q--average of quadratic invariant scalars ($E_{ab}E^{ab},\,\sigma_{ab}E^{ab},\,\sigma_{ab}\sigma^{ab}$) that vanish for FLRW models.

It is important to remark that the notion of a gravitational entropy connected to inhomogeneity (marked by invariant scalars of the Weyl tensor) bears a close relation to the concept of the ``arrow of time'' originally suggested by Penrose \cite{arrow1}, and further explored by a number of authors \cite{arrow2,arrow3,arrow4,arrow5,arrow6}. The Weyl tensor scalar that emerges in this literature is $C_{abcd}C^{abcd}$ (proportional to $E_{ab}E^{ab}$ in LTB models) and no attempt is made in these papers to connect it with averages and fluctuations.  As a contrast, the Leibler--Kullback functional is based on an averaging formalism that yields a direct connection between the growth of structure (inhomogeneity) and its associated fluctuations, which in the case of the q--average, are directly related to invariant contractions of $\sigma_{ab}$ and $ E^{ab}$, as well as with fluctuations and statistical variance and covariance moments of both $\rho$ and $\HH$ (which is a fluid velocity). Evidently, we have only considered this functional with the q--average applied on LTB models, and thus, we cannot exclude the possibility that contradictory results may happen with Buchert's average or in other types of models (as for example with the ``arrow of time'', see \cite{arrow4}), but the work we have developed here may provide a guideline on how to explore the notion of the arrow of time and its relation with the Leibler--Kullback functional in  more general models.

While the q--average formalism cannot be used to study and understand the dynamical implications of back--reaction, it has resulted very helpful for understanding the properties and evolution of LTB models \cite{sussDS1,sussDS2,RadAs,RadProfs}, including the computation in these models of back--reaction terms and verification of the existence of ``effective'' acceleration in the context of Buchert's formalism \cite{suss2011}. Moreover, it does provide through its fluctuations and perturbations interesting connections between averaging, perturbation theory, invariant scalars and statistical correlations of $\rho$ and $\HH$, which signals a valuable theoretical insight on how the averaging process should work in any generic solution of Einstein's equations (at least in LRS spacetimes whose dynamics is reducible to scalar modes). In particular, since most formal and theoretical results obtained for LTB models can be readily applied to Szekeres models \cite{sussbol,peel}, the extension of our results to these models is currently under elaboration. In fact, the dynamical effect of q--scalars in LTB models may be applicable to even more general spacetimes. We feel that exploring this proposal is worth considering in future research. 

\section*{Acknowledgements.} 

The author acknowledges financial support from grant SEP--CONACYT 132132. Acknowledgement is also due to Thomas Buchert for useful comments.           

\begin{appendix}

\section{Analytic solutions.}

The metric function $R$ in (\ref{ltb}) can be obtained from the following parametric solutions of (\ref{fieldeq1}) which define parabolic, hyperbolic and elliptic models:
\ba 
\fl \hbox{\bf{Parabolic}:}\;\;E=0,\quad R=(2M)^{1/3}\,\eta^2,\qquad t-\tbb = \frac{2}{3}\,\eta^3,\nonumber\\
\fl \hbox{\bf{Hyperbolic}:}\;\;E\geq 0,\quad R =\frac{M}{2E}\,\left(\cosh\,\eta-1\right),\quad 
t-\tbb=\frac{M}{(2E)^{3/2}}\,\left(\sinh\,\eta-\eta\right),\nonumber\\
\fl \hbox{\bf{Elliptic}:}\;\;-1<E\leq 0,\quad R =\frac{M}{2|E|}\,\left(1-\cos\,\eta\right),\quad
t-\tbb =\frac{M}{(2|E|)^{3/2}}\,\left(\eta-\sin\,\eta\right),\nonumber\\
\label{ellt1}
\ea
where $\tbb=\tbb( r)$, which emerges as an ``integration constant'', is the  ``big bang time'' because it  marks the coordinate locus of the central expanding curvature singularity $R(t,r)=0$ for $r\ne$ constant. The classification of (\ref{ellt1}) follows from the sign of $E$ and governs the kinematic evolution of dust layers through the existence of a zero of $\dot R$ in (\ref{fieldeq1}): parabolic and hyperbolic layers  perpetually expand and elliptic layers expand/collapse. Since $E=E(r)$, the models can exhibit, either the same kinematic pattern for all $r$, or ``mixed'' patterns that define models with regions of different kinematic behavior as $E$ changes sign in specific ranges of $r$ (see examples in \cite{kras1,kras2,KH1,KH2,KH3,BoKrHe,ltbstuff}).

The solutions (\ref{ellt1}) allow for a complete description of the dynamics of the models once we select the free functions $M,\,E,\,\tbb$, though only two of these functions are strictly necessary because the metric (\ref{ltb}) in invariant under re--scalings $r=r(\bar r)$ and thus any one of these free functions can be eliminated by a suitable choice of the radial coordinate. In particular, the expressions (\ref{fieldeq2}), (\ref{HH}), (\ref{KK}), (\ref{Sigma1}) and (\ref{EE1}) for the basic covariant scalars $\rho,\HH,\,\KK,\,\Sigma,\,\EE$ in (\ref{locscals}) involve terms such as $R'$ and $\dot R'$, which can be computed by implicit derivation of (\ref{ellt1}) with respect to $r$.  

\section{Functions of q--scalars are q--scalars.}

The relations (\ref{rder}) allow us to express local scalars $A$ in terms of q--scalars $A_q$ through the differential inverse of the integral definition (\ref{qave}):
\begin{equation} A = \Aav_q +\frac{\Aav'_q}{\VV_q'/\VV_q}=A_q+\frac{A'_q}{\VV_q'/\VV_q},\label{invdef}
\end{equation}
where $\VV_q'/\VV_q=3R'/R$ and $\VV_q$ is defined in (\ref{qvol}). As a consequence, any scalar expressible as a function of q--scalars is itself a q--scalar: if $A_q$ and $B_q$ are two q--scalars and $U=U(A_q,B_q)$, then
\begin{equation} U' = \frac{3R'}{R}\left[\frac{\partial U}{\partial A_q}(A-A_q)+\frac{\partial U}{\partial B_q}(B-B_q)\right],\label{fprime}\end{equation}
so that we can identify $Z_q = U(A_q,B_q)$ as the q--scalar whose corresponding ``local'' scalar $Z$ is given by
\begin{equation} Z = Z_q +\frac{\partial U}{\partial A_q}(A-A_q)+\frac{\partial U}{\partial B_q}(B-B_q)=Z_q+\frac{Z'_q}{3R'/R},\label{eqZ}\end{equation}
where it is important to remark that (in general) we have $Z\ne U(A,B)$. It is straightforward to show that $Z$ and $Z_q$ above satisfy the integral definition (\ref{qave}), which basically states that a function of q--averages $U(\Aav_q,\Bav_q)$ is itself the q--average $\langle Z\rangle_q$ of a scalar $Z$ given by (\ref{eqZ}), which in general does not coincide with $U(A,B)$ nor its q--average $\langle U(A,B)\rangle_q$.

\section{Proof of (\ref{var}) and (\ref{cov}).}

The proof of (\ref{var}) for $A=\HH$  is equivalent to the proof that the identity $\langle W_\HH(r,r_0)\rangle_q[r_b]=0$ holds for every domain $\DD[r_b]$, where the scalar $W_\HH$ is given by:
\begin{equation}\fl W_\HH(r,r_b) = [\DENL(\HH)]^2-[\Del(\HH)]^2=\left(\,\HH( r)-\HHav_q[r_b]\,\right)^2-\left(\,\HH( r)-\HH_q( r)\,\right)^2.\label{B11}\end{equation}
We expand $W_\HH(r,r_b)$ and apply (\ref{aveave}), so that $\langle -2\HH( r)\HHav_q[r_b]+\HHav_q^2[r_b]\rangle_q[r_b]=-\HHav_q^2[r_b]$ holds, leading to 
\begin{equation} \fl \langle W_\HH(r,r_b)\rangle_q[r_b]=-\HHav_q^2[r_b]+\frac{1}{\VV_q(r_b)}\int_0^{r_b}{(2\HH\HH_q-\HH_q^2)\,\VV'_q\,\dd\bar r}.\label{B12}\end{equation}
where $\VV_q$ is defined by (\ref{qvol}), so that $\VV'_q = 4\pi R^2 R'$. Considering that (\ref{HH}) and (\ref{rhoHKq}) imply $3\HH=\theta=\dot\VV'_q/\VV'_q$ and $3\HH_q=\theta_q=\dot\VV_q/\VV_q$. Inserting these identities into (\ref{B12}) yields the desired result:
\ba \fl \langle W_\HH(r,r_b)\rangle_q[r_b] &=& -\HHav_q^2[r_b]+\frac{1}{\VV_q(r_b)}\int_0^{r_b}{\left(\frac{\dot\VV_q^2}{9\VV_q}\right)'\,\dd\bar r}=-\HHav_q^2[r_b]+\HH_q^2( r_b)=0,\nonumber\\
\label{B13}\ea
since $\HHav_q[r_b]$ coincides with $\HH_q$ at the domain boundary $r=r_b$. 

For $A=\rho$ the proof of (\ref{var}) is equivalent to proving $W_\rho(r,r_b) \rangle_q[r_b]=0$, where 
\begin{equation} \fl W_\rho(r,r_b) =  [\DENL(\rho)]^2-[\Del(\rho)]^2=\left(\,\rho( r)-\rhoav_q[r_b]\,\right)^2-\left(\,\rho( r)-\rho_q( r)\,\right)^2.\label{B21}\end{equation}
Expanding $W_\rho(r,r_b)$, applying (\ref{aveave}) and substituting $\rho = M'/\VV'_q$ and $\rho_q = M/\VV_q$ from (\ref{fieldeq2}) and (\ref{rhoHKq}), yields a relation similar to (\ref{B12}): 
\ba \fl \langle W_\rho(r,r_b)\rangle_q[r_b] &=& -\rhoav_q^2[r_b]+\frac{1}{\VV_q(r_b)}\int_0^{r_b}{(2\rho\rho_q-\rho_q^2)\,\VV'_q\,\dd\bar r},\nonumber\\
\fl &=& -\rhoav_q^2[r_b]+\frac{1}{\VV_q(r_b)}\int_0^{r_b}{\left(\frac{M^2}{\VV_q}\right)'\,\dd\bar r}=-\rhoav_q^2[r_b]+\rho_q^2( r_b)=0.\nonumber\\
\label{B23}\ea
The proof of (\ref{cov}) for $A,B=\rho,\,\HH$ follows along similar lines: it is equivalent to proving that: $\langle\,W_{\rho\HH}(r,r_b)\, \rangle_q[r_b]=0$ holds, where:
\ba \fl W_{\rho\HH}(r,r_b) = \DENL(\HH)\,\DENL(\rho)-\Del(\HH)\,\Del(\HH)\nonumber\\ 
\fl = \left(\,\rho( r)-\rhoav_q[r_b]\,\right)\left(\,\HH( r)-\HHav_q[r_b]\,\right)-\left(\,\rho( r)-\rho_q( r)\,\right)\left(\,\HH( r)-\HH_q( r)\,\right).
\label{B31}\ea
Substitution of the forms for $\HH,\,\HH_,\,\rho,\,\rho_q$ given before yields the desired result:
\ba \fl \langle W_{\rho\HH}(r,r_b)\rangle_q[r_b] 
=-\rhoav_q[r_b]\HHav_q[r_0]+\frac{1}{\VV_q(r_b)}\int_0^{r_b}{(\rho\HH_q+\rho_q\HH-\rho_q\HH_q)\,\VV'_q\,\dd\bar r}\nonumber\\
=-\rhoav_q[r_b]\HHav_q[r_b]+\frac{1}{\VV_q(r_b)}\int_0^{r_b}{\left(\frac{M\dot\VV_q}{3\VV_q}\right)'\,\dd\bar r}\nonumber\\
=-\rhoav_q[r_b]\HHav_q[r_b]+\rho_q( r_b)\HH_q( r_b)=0.
\label{B32}\ea
\section{Proper curvature and kinematic tensors.}

We show in this appendix that all relevant (curvature and kinematic) proper tensors characteristic of LTB models are completely determined by 4 invariant scalars: the Ricci scalar,  $\CR=\CR^c_c$, the only nonzero Newman--Penrose conformal invariant, $\Psi_2$, the Hubble expansion scalar $\HH=\theta/3$ and the eigenvalue of the shear tensor, $\Sigma$.  

The Riemann and Weyl tensors are given by the following expressions:
\begin{equation} \fl \CR^{ab}_{cd}=C^{ab}_{cd}-\frac{1}{2}\CR\delta^a_{[c}\delta^b_{d]}+2\delta^{[a}_{[c}\CR^{b]}_{d]},\qquad 
C^{ab}_{cd} = \delta^{[a}_{[c}E^{b]}_{d]}+2 u_{[c}E^{[a}_{d]}u^{b]},\label{RieWeyl}\end{equation}
where square bracketed indices indicate anti--simetrization. The Ricci and electric Weyl tensors, $\CR^b_d$ and $E^b_d$ follow from contraction of the tensors in (\ref{RieWeyl}):
\ba \fl\CR^b_d = \CR^{bc}_{cd}=\frac{1}{2}\CR\left(h^b_d+u^b u_d\right),\qquad E^b_d = u_au_c C_{d}^{abc}=\Psi_2 \,\hbox{\bf{e}}^b_d,\label{RE}\ea
where $\hbox{\bf{e}}^b_d=\delta^b_d-3n^bn_d$ is the common eigenframe for $E^b_d$ and the shear tensor $\sigma^b_d$ defined in (\ref{Sigma1}) and (\ref{EE1}).  By combining (\ref{RieWeyl}) and (\ref{RE}) we get
\ba \fl \CR^{ab}_{cd}=C^{ab}_{cd}+\frac{1}{3}\CR\left(3\delta^{[a}_{[c}\delta^{b]}_{d]}+6\delta^{[a}_{[c}u^{b]}u_{d]}-\delta^a_{[c}\delta^b_{d]}\right),\label{Rie2}\\
\fl C^{ab}_{cd} = \Psi_2\,\left(\delta^{[a}_{[c}\hbox{\bf{e}}^{b]}_{d]}+2u_{[c}\hbox{\bf{e}}^{[a}_{d]}u^{b]}\right)=\Psi_2\,\left(h^{[a}_{[c}-3u_{[c}u^{[a}\right)\,\hbox{\bf{e}}^{b]}_{d]},\label{Weyl2}\ea
Hence, all curvature proper tensors are determined by two invariant scalars: $\CR$ and $\Psi_2$, which relate to the density and its local fluctuation by $\CR = 8\pi\rho$ and $\Psi_2=-\EE=(4\pi/3)(\rho-\rho_q)=(4\pi/3)\Del(\rho)$. 

The kinematic proper tensors are the shear and ``expansion'' tensors: $\sigma_{ab},\,\HH_{ab}$. Both are expressible in terms of the invariant scalars $\HH=\theta/3=\tilde\nabla_au^a/3$ and $\Sigma = \sigma_{ab}\hbox{\bf{e}}^{ab}$: 
\begin{equation} \sigma_{ab} = \Sigma\,\hbox{\bf{e}}_{ab},\qquad \HH_{ab}=\HH h_{ab}+\Sigma\,\hbox{\bf{e}}_{ab}, \end{equation}
where the explicit forms of $\HH$ and $\Sigma$ are given in (\ref{HH}) and (\ref{Sigma1}). Notice that $\Sigma=-\Del(\HH)$ is a local fluctuation of $\HH$.     

\end{appendix}


\section*{References}

\begin{thebibliography}{9}


\bibitem{Lemaitre} Lema\^{\i}tre G  1933 {\it Ann. Soc. Sci. Brux.} A  {\bf 53} 51. See reprint in Lema\^{\i}tre G  1997 {\it Gen. Rel. Grav.} {\bf 29} 5.

\bibitem{TB} Tolman R C 1934 {\it Proc. Natl Acad. Sci.} {\bf 20} 169; Sen N R 1934 {\it Z Astrophysik} {\bf 9} 215 (see reprint in Sen N R 1997 {\it Gen Rel Gravit} {\bf 29} 1473); Bondi H 1947 {\it Mon. Not. R. Astron. Soc.} {\bf 107} 410.


\bibitem{kras1} Krasi\'nski A, {\textit{Inhomogeneous Cosmological Models}},  Cambridge University Press, 1998.

\bibitem{kras2} Plebanski J and Krasinski A, {\textit{An Introduction to General Relativity and Cosmology}},  Cambridge University Press, 2006.

\bibitem{BKHC2009} K. Bolejko, A. Krasi\'nski, C. Hellaby, M.-N. C\'el\'erier, Structures in the Universe by exact methods: formation, evolution, interactions, Cambridge University Press, Cambridge 2009

\bibitem{celerier} Celeri\`er M N 2007 {\it New Advances in Physics} {\bf 1} 29 ({\it Preprint} {\tt arXiv:astro-ph/0702416})

\bibitem{KH1} Krasi\'nski A and Hellaby C 2002 {\it Phys Rev} D {\bf 65} 023501

\bibitem{KH2} Krasi\'nski A and Hellaby C 2004 {\it Phys Rev} D {\bf 69} 023502

\bibitem{KH3} Krasi\'nski A and Hellaby C 2004 {\it Phys Rev} D {\bf 69} 043502

\bibitem{KH4} Hellaby C and Krasi\'nski A 2006 {\it Phys Rev} D {\bf 73} 023518

\bibitem{BoKrHe} Bolejko K Krasi\'nski A and Hellaby C 2005 {\it MNRAS} {\bf 362} 213 ({\it Preprint} {\tt arXiv:gr-qc/0411126})

\bibitem{ltbstuff}  Matravers D R and Humphreys N P  2001 {\it Gen. Rel. Grav.} {\bf 33} 
531Ð52; Humphreys N P, Maartens R and Matravers D R 1998 Regular spherical dust spacetimes ({\it Preprint} {\tt gr-qc/9804023v1})

\bibitem{focus} Bolejko K Celerier M N and Krasinski A 2011 {\it Class. Quant. Grav.} {\bf 28} 164002; 


\bibitem{obs1}  Pascual--S\'anchez J F 1999  {\it Mod. Phys. Lett.} A {\bf 14} 1539;  Sugiura N K and Harada T 1999  {\it Phys Rev } D 60 103508; Celeri\`er M N 2000 {\it Astron. Astrophys.} {\bf 353} 63;   Tomita K 2001 {\it MNRAS} {\bf 326} 287;  Iguchi H,  Nakamura T and Nakao K 2002 {\it Prog. Theor. Phys.} {\bf 108} 809; Schwarz D J  2002 Accelerated expansion without dark energy ({\it Preprint} {\tt arXiv:astro-ph/0209584v2});    

\bibitem{obs2}    Apostolopoulos P {\it et al}  2006 {\it JCAP} {\bf P06} 009; Kai T, Kozaki H, Nakao K, Nambu Y and Yoo C M 2007 {\it Prog. Theor. Phys.} {\bf 117} 229-240 ({\it Preprint} {\tt  arXiv:gr-qc/0605120});   Mattsson T and Ronkainen M 2008  {\it JCAP} {\bf 0802} 004 ({\it Preprint} {\tt arXiv:astro-ph/0708.3673v2});   Bolejko K and Andersson L 2008 {\it JCAP} {\bf 10} 003 ({\it Preprint} {\tt arXiv:0807.3577});  Rasanen S 2006 {\it Class. Quant. Grav.} {\bf 23} 1823-1835; Moffat J W 2006 {\it JCAP}05(2006)001 

\bibitem{kolb} Kolb E W, Matarrese S, Notari A and  Riotto A 2005 {\it Phys Rev} D {\bf 71} 023524  ({\it Preprint} {\tt arXiv:hep-ph/0409038v2}); Marra V,  Kolb E W and Matarrese S 2008 {\it Phys Rev} D {\bf 77} 023003; Marra V, Kolb E W, Matarrese S and Riotto A 2007 {\it Phys Rev} D {\bf 76} 123004. 

\bibitem{GBH} Garc\'\i a--Bellido J and Troels H 2008 {\bf JCAP} 0804:003 ({\it Preprint} {\tt gr-qc/0802.1523v3 [astro-ph]}) 

\bibitem{alnes} Alnes H, Amazguioui M and Gron O 2006 {\it Phys Rev} D {\bf 73} 083519; Alnes H and Amazguioui M 2006 {\it Phys Rev} D {\bf 74} 103520;  Alnes H and Amazguioui M 2006 {\it Phys Rev} D {\bf 75} 023506

\bibitem{bisetal} Biswas T Mansouri R and Notari A 2007 {\it JCAP} 0712:017,
{\tt arXiv:astro-ph/0606703v2}; Biswas T and Notari A 2008 {\it JCAP} 0806:021 {\tt arXiv:astro-ph/0702555}; Alexander S {\it et al} 2009 {\it JCAP} 0909:025 {\it arXiv:0712.0370}; Biswas T Notari A and Valkenburg W 2010 {\it JCAP} {\bf 11} 030 ({\it Preprint} {\tt arXiv:1007.3065})  

\bibitem{endqvist} Enqvist K and  Mattsson T 2007 {\it JCAP} {\bf 0702} 019 ({\it Preprint} {\tt 	arXiv:astro-ph/0609120v4}); Enqvist K 2008 {\it Gen. Rel. Grav.}  {\bf 40}  451-466 ({\it Preprint} {\tt arXiv:0709.2044})  

\bibitem{clarkson} February S {\it et al} 2010 {\it MNRAS} {\bf 405} 2231--2242 ({\it Preprint} {\tt arXiv:0909.1479v2[astro-ph CO]})

\bibitem{kras3} C\'el\'erier M N  Bolejko K and Krasi\'nski A 2010 {\it Astron Astrophys} {\bf 518} A21 {\tt  arXiv:0906.0905} 

\bibitem{marranot} Marra V and Notari A 2011 {\it Class. Quant. Grav.} {\bf 28} 164004 ({\it Preprint} {\tt arXiv:1102.1015})


\bibitem{lemos} Eardley D M 1974 {\it Commun Math Phys} {\bf 37} 287; Eardley D M and Smarr L 1979 {\it Phys Rev} D {\bf 19} 2239; Dyer C C 1979 {\it MNRAS} {\bf 189} 189; Waugh B and Lake K 1988 {\it Phys Rev} D {\bf 38} 1315; Waugh B and Lake K 1989 {\it Phys Rev} D {\bf 40} 2137; Lemos J P S 1991 {\it Phys Lett} A {\bf 158} 279

\bibitem{joshi} Joshi P S and Dwivedi I H 1993 {\it Phys Rev} D {\bf 47} 5357; Joshi P S and Singh T P 1995 {\it Phys Rev} D {\bf 51} 6778; Dwivedi I H and Joshi P S 1997 {\it Class. Quant. Grav.} {\bf 47} 5357


\bibitem{quantum} Vaz C, Witten L and Singh T P 2001 {\it Phys Rev} D {\bf 63} 104020;
Kiefer C, Mueller-Hill, Vaz C 2006 {\it Phys Rev} D {\bf 73} 044025;
Bojowald M, Harada T and Tibrewala R 2008 {\it Phys Rev} D {\bf 78} 064057




\bibitem{ellisbruni89} Ellis G F R and Bruni M 1989 {\it Phys Rev} D {\bf 40}  1804

\bibitem{BDE} Bruni M,  Dunsby P K S and Ellis G F R 1992 {\it Astroph. J.} {\bf 395} 34--53

\bibitem{1plus3} Ellis G F R and van Elst H 1998 Cosmological Models (Carg\`ese Lectures 1998) {\it Preprint} {\tt arXiv gr-qc/9812046 v4}

\bibitem{zibin}  Zibin J P 2008 {\it Phys Rev} D{\bf 78} 043504 [arXiv:0804.1787]

\bibitem{dunsbyetal} Dunsby P {\it et al} ({\it Preprint} {\tt  arXiv:1002.2397v1 [astro-ph CO]})

\bibitem{LRS} van Elst H and  Ellis G F R 1996 {\it Class Quantum Grav} {\bf 13} 1099-1128 ({\it Preprint} {\tt  arXiv:gr-qc/9510044})


\bibitem{wainwright} Wainwright J and Andrews S 2009 {\it Class.Quant.Grav.},{\bf 26}, 085017


\bibitem{buchert} Buchert T, 2000 {\it  Gen. Rel. Grav.} {\bf 32} 105; Buchert T, 2000 {\it Gen.Rel.Grav.} {\bf 32} 306-321; Buchert T 2001 {\it Gen. Rel. Grav.} {\bf 33} 1381-1405;    Ellis G F R and Buchert T 2005 {\it Phys.Lett.} A{\bf 347} 38-46; Buchert T and Carfora M 2002 {\it Class.Quant.Grav.} {\bf 19} 6109-6145;  Buchert T 2006 {\it Class. Quantum Grav.} 23 819;  Buchert T, Larena J and Alimi J M 2006 {\it Class. Quantum Grav.} 23 6379;  Buchert T 2005 {\it Class. Quantum Grav.} {\bf 22} L113--L119;  Buchert T 2006 {\it Class. Quantum Grav.} {\bf 23} 817--844 ({\it Preprint} {\tt arXiv:gr-qc/0509124})

\bibitem{buchGRG} Buchert T 2008 {\it Gen. Rel. Grav.}  {\bf 40}  467--527 ({\it Preprint} {\tt arXiv:xxxx.xxxx})  

\bibitem{buchCQG} Buchert T 2011 {\it Class. Quantum Grav.} {\bf 28} 164007 ({\it Preprint} {\tt arXiv:1103.2016})


\bibitem{zala} Zalaletdinov R M, {\it Averaging Problem in Cosmology and
Macroscopic Gravity}, Online Proceedings of the Atlantic Regional Meeting on General Relativity and Gravitation, Fredericton, NB, Canada, May 2006, ed. R.J. McKellar ({\it Preprint}  {\tt arXiv:gr-qc/0701116})

\bibitem{colpel} Coley A A and Pelavas N 2007 {Phys.Rev.} D {\bf 75} 043506; Coley A A, Pelavas N and Zalaletdinov R M 2005 {\it Phys.Rev.Lett.} {\bf 95} 151102


\bibitem{avedebate} Buchert T and Carfora M 2002 {\it Class. Quantum Grav.} {\bf 19} ({\it Preprint} {\tt arXiv:gr-qc/0210037});  Paranjape A and Singh T P 2008 {\it Gen.Rel.Grav.} {\bf 40} 139-157 ({\it Preprint} {\tt arXiv:astro-ph/0609481v4});  Paranjape A and Singh T P 2007 {\it Phys.Rev.} D {\bf 76} 044006 ({\it Preprint} {\tt arXiv:gr-qc/0703106v3});  Paranjape A 2008 {\it Int.J.Mod.Phys.} D {\bf 17} 597-601 ({\it Preprint} {\tt arXiv:0705.2380v1 [gr-qc]}); Marozzi G 2011 {\it JCAP} {\bf 01} 012 ({\it Preprint} {\tt arXiv:1011.4921 [gr-qc]})


\bibitem{sussQL} Sussman R A Quasi-local variables and inhomogeneous cosmological sources with spherical symmetry 2008 {\it AIP Conf.Proc.} {\bf 1083} 228-235 {\tt Preprint arXiv:0810.1120}.

\bibitem{suss2009} Sussman R A 2009 {\it Phys Rev} D {\bf 79} 025009 ({\it Preprint} {\tt arXiv:arXiv:0801.3324 [gr-qc]})


\bibitem{sussbol} Sussman R A and Bolejko K 2012 {\it Class Quant Grav} {\bf 29} 065018 ({\it Preprint} {\tt ArXiv 1109.1178})

\bibitem{peel} Peel A Ishak M and Troxel M A 2012 {\it Phys.Rev.} D {\bf 86} 123508 ({\it Preprint} {\tt arXiv:1212.2298});


\bibitem{buchstruct} Buchert T 2011 {\it Classical and Quantum Gravity} {\bf 28} 164007;  Clarkson C  Ellis G F R  Larena J and  Umeh O 2011 {\it Reports on Progress in Physics}, ({\it Preprint} {\tt arXiv:1109.2314}); Clarkson C Ananda K and Larena J 2009 {\it Phys Rev} D {\bf 80} 083525 ({\it Preprint} {\tt arXiv:0907.3377}); Ras\"anen S 2008 {\it JCAP} {\bf 0804} 027 ({\it Preprint} {\tt arXiv:0801.2692}); Paranjape A 2008 {\it Phys.Rev.D} {\bf 78} 063522 ({\it Preprint} {\tt arXiv:0806.2755v2 [astro-ph]});  Paranjape A and Singh T P 2008 {\it Phys.Rev.Lett.} {\bf 101} 181101 ({\it Preprint} {\tt arXiv:0806.3497v3 [astro-ph]}); Mattsson T and Ronkainen M 2008 {\it JCAP} {\bf 02} 004;  Buchert T 2006 {\it Astron Astrophys} {\bf 454} 415-422 ({\it Preprint} {\tt arXiv:astro-ph/0601513})


\bibitem{buchobs} Kolb E 2011 {\it Classical and Quantum Gravity} {\bf 28} 164009; Clarkson C and Umeh O 2011 {\it Classical and Quantum Gravity} {\bf xx} ({\it Preprint} {\tt arXiv:1105.1886})   Larena J Alimi J M Buchert T Kunz M and Corasaniti P S 2009 {\it Phys.Rev.} D {\bf 79} 083011 ({\it Preprint} {\tt arXiv:0808.1161});  Kolb E W Matarrese S and Riotto A 2006 {\it NewJ.Phys.} {\bf 8} 322 ({\it Preprint} {\tt arXiv:astro-ph/0506534});


\bibitem{buchacc} Buchert T 2005 {\it Class Quant Grav} {\bf 22} L113-L119 ({\it Preprint} {\tt arXiv:gr-qc/0507028});  Ellis GFR and Buchert T 2005 {\it Phys Lett} A {\bf 347} 38-46 ({\it Preprint} {\tt arXiv:gr-qc/0506106}); R\"a\"sanen S 2011 {\it Class Quantum Grav} {\bf 28} 164008 ({\it Preprint} {\tt arXiv:1102.0408})



\bibitem{buchdark} Buchert T and Obadia N 2011 {\it Class Quant Grav} {\bf 28} 162002 ({\it Preprint} {\tt arXiv:1010.4512}); Wiegand A and Buchert T 2010 {\it Phys Rev} D{\bf 82} 023523 ({\it Preprint} {\tt arXiv:1002.3912}); Roy X and Buchert T 2010 {\it Class Quant Grav} {\bf 27} 175013   ({\it Preprint} {\tt arXiv:0909.4155})


\bibitem{LTBave1} Chuang C H, Gu J A and Hwang W Y P 2005 {\it Class.Quant.Grav.},{\bf 25}, 175001 {\tt Preprint  astro-ph/0512651}

\bibitem{LTBave2} Paranjape A and Singh T P 2006 {\it Class Quant Grav} {\bf 23}, 6955Ð6969

\bibitem{mattsson} Mattsson M and Mattsson T 2010 {\it JCAP} {\bf 10} 021; Mattsson M and Mattsson T 2011 {\it JCAP} {\bf 05} 003;

\bibitem{sussBR} Sussman R A 2008 On spatial volume averaging in Lema"tre--Tolman--Bondi dust models. Part I: back reaction, spacial curvature and binding energy  ({\it Preprint} {\tt arXiv:0807.1145 [gr-qc]})

\bibitem{sussIU} Sussman R A 2010 {\it AIP Conf Proc} {\bf 1241} 1146-1155 ({\it Preprint} {\tt arXiv:0912.4074 [gr-qc]})

\bibitem{suss2011} Sussman R A 2011 {\it Class Quantum Grav} {\bf 28} 235002 ({\it Preprint} {\tt arXiv:1102.2663v1 [gr-qc]})


\bibitem{sussDS1}Sussman R A 2008 {\it Class Quantum Grav.}  {\bf 25} 015012 {\tt Preprint arXiv:grÐqc/0709.1005}

\bibitem{sussDS2}Sussman R A and Izquierdo G 2011 {\it Class Quantum Grav.}  {\bf 28} 045006 {\tt Preprint arXiv:grÐqc/1004.0773}

\bibitem{RadAs} Sussman R A 2010 {\textit{Gen Rel Grav}} {\bf{42}} 2813--2864 ({\it Preprint} {\tt arXiv:1002.0173 [gr-qc]})

\bibitem{RadProfs} Sussman R A 2010 {\textit{Class.Quant.Grav.}} \textbf{27} 175001 ({\it Preprint} {\tt arXiv:1005.0717 [gr-qc]})


\bibitem{entropy1} Hosoya A Buchert T and Morita M 2004 {\it Phys Rev Lett} {\bf 92}, 141302-1Ð4 ({\it Preprint} {\tt arXiv:gr-qc/0402076})

\bibitem{entropy2}
Morita M Buchert T Hosoya A and Li N 2010  {\it AIP Conf.Proc.} {\bf 1241} 1074-1082 ({\it Preprint} {\tt arXiv:1011.5604})

\bibitem{entropy3} Li N Buchert T Hosoya A Morita M and Schwarz D J 2012 ``Relative information entropy and Weyl curvature of the inhomogeneous Universe'' ({\it Preprint} {\tt arXiv:1208.3376})
 
 arXiv:1208.3376



\bibitem{podurets} Podurets M A 1964 {\it Sov Astr A J} {\bf 8} 19. 

\bibitem{MiSh} Misner C and Sharp D H 1964 {\it Phys Rev} B {\bf 136} 571

\bibitem{hayward} Hayward S A 1996 {\it Phys Rev} D {\bf 53} 1938 ({\it Preprint} {\tt ArXiv gr-qc/9408002}); Hayward S A 1998 {\it Class Quantum Grav} {\bf 15} 3147Ð3162  ({\it Preprint} {\tt ArXiv gr-qc/9710089v2})


\bibitem{stephani} Stephani H 1967 {\it Commun Math Phys} {\bf 4} 137


\bibitem{arrow1} Penrose R 1979 in {\it General Relativity, an Einstein Centenary Survey}. Edited by Hawking S W and Israel W, Cambridge University Press.

\bibitem{arrow2} Wainwright J 1984 {\textit{Gen Rel Grav}} {\bf{16}} 657

\bibitem{arrow3} Bonnor W B 1986 {\it Class Quantum Grav} {\bf 3} 495 

\bibitem{arrow4} Bonnor W B 1987 {\it Phys Lett} A {\bf 122} 305

\bibitem{arrow5} Gron O and Hervik S 2002 ``The Weyl curvature conjecture'' ({\it Preprint}  {\tt gr-qc/0205026v1})

\bibitem{arrow6} Pelavas N and Lake K 2000 {\it Phys Rev} D {\bf 62} 044009





\end{thebibliography}
\end{document}